# Thermal Transport in Ag$_8$TS$_6$ (*T*= Si, Ge, Sn) Argyrodites: An Integrated Experimental, Quantum-Chemical, and Computational Modelling Study

Joana Bustamante,[a] Anupama Ghata,[b] Aakash A. Naik,[a,c] Christina Ertural,[a] Katharina Ueltzen,[a,c] Wolfgang G. Zeier,[b,d] and Janine George *[a,c]

Argyrodite-type Ag-based sulfides combine exceptionally low lattice thermal and high ionic conductivity, making them promising candidates for thermoelectric and solid-state energy applications. In this work, we studied Ag$_8$TS$_6$ (*T*= Si, Ge, Sn) argyrodite family by combining chemical-bonding analysis, lattice vibrational properties simulation, and experimental measurements to investigate their structural and thermal transport properties. Furthermore, we propose a two-channel lattice-dynamics model based on Grüneisen-derived phonon lifetimes and compare it to an approach using machine-learned interatomic potentials. Both approaches are able to predict thermal conductivity in agreement with experimental lattice thermal conductivities along the whole temperature range, highlighting their potential suitability for future high-throughput predictions. Our findings also reveal a relationship between bond heterogeneity arising from weakly bonded Ag$^+$ ions and occupied antibonding states in Ag–S and Ag–Ag interactions and strong anharmonicity including large Grüneisen parameters, and low sound velocities, which are responsible for the low lattice thermal conductivity of Ag$_8$SnS$_6$, Ag$_8$GeS$_6$, and Ag$_8$SiS$_6$. We furthermore show that thermal and ionic conductivities in all three compounds are independent of each other and can likely be tuned individually.

## Introduction

To reduce the enormous waste of heat in energy generation, thermoelectric materials (TE) offer a promising solution for energy saving and environmental protection. They can convert heat into electricity or vice versa. The thermal conductivity of a material is crucial for its thermoelectric efficiency and a lower thermal conductivity results in higher efficiency. For example, several argyrodites such as Ag$_8$GeSe$_6$ and Cu$_7$PSe$_6$ and the isovalently substituted compounds Ag$_8$SiSe$_6$ and Ag$_8$SnSe$_6$ are known for their high ionic conductivity and many others have been investigated as potential thermoelectrics.[1–8] Halogen-free argyrodites have a general chemical formula of $A^{m+}_{(12-n)/m}T^{n+}Q^{2-}_6$ (A=Ag, Cu; T=Si, Ge, Sn; and Q=S, Se and Te).[2,8–11] While high ionic conductivity could be problematic for the stability of a thermoelectric device, we and others have demonstrated that thermal and ionic conductivity of some Ag$^+$ and Cu$^+$ based argyrodites (e.g., Ag$_8$GeSe$_6$, Ag$_{8-x}$Cu$_x$GeS$_6$, and Cu$_7$PSe$_6$) are not directly correlated with each other and can also be tuned independently.[8,12–14] A similar situation might be expected for the canfieldite Ag$_8$SnS$_6$ and Ag$_8$SiS$_6$, which are argyrodite family members and isovalently substituted variants of Ag$_8$GeS$_6$.

The canfieldite (Ag$_8$SnS$_6$) shows promising thermoelectric (TE) properties. Shen and co-workers evaluated its lattice thermal conductivity and its crystal structure in detail, finding an orthorhombic $Pna2_1$ crystal structure at room temperature.[15] Slade's study reported an additional orthorhombic $Pmn2_1$ phase at 120 K, also indicating potential TE properties.[16] Additionally, a previous study pointed out the importance of thermal transport via a diffusive transport mechanism.[17] All previous studies suggested that the weakly bonded Ag$^+$ ions contribute to the low lattice thermal conductivity in the canfieldites Ag$_8$TS$_6$ (*T* = Si, Ge, Sn).[13,15,16] However, a complete understanding of the connection between lattice thermal conductivity, ionic conductivity, and their correlation with bonding, anharmonicity, and elastic properties remains unexplored for all three compounds.

Several models have been developed to estimate lattice thermal conductivity with limited computational resources, each offering varying degrees of mathematical complexity and accuracy. However, no existing model is both computationally efficient enough for high-throughput studies and reliably accurate across the entire temperature range. Traditional models such as Slack[18–20] take into account the importance of acoustic phonons and elastic properties, often providing a temperature-dependent lattice thermal

a. Federal Institute for Materials Research and Testing (BAM), Department of Materials Chemistry, Unter den Eichen 87, 12205 Berlin, Germany.
b. University of Münster, Institute of Inorganic and Analytical Chemistry, Corrensstr. 28/30, D-48149 Münster, Germany.
c. Friedrich Schiller University Jena, Institute of Condensed Matter Theory and Solid-State Optics, Max-Wien-Platz 1, 07743 Jena, Germany.
d. Institute of Energy Materials and Devices (IMD), IMD-4: Helmholtz-Institut Muenster, Forschungszentrum Juelich, 48149 Münster, Germany.

conductivity ($\kappa_L$), but occasionally with overestimated values. The Cahill[21] and Agne[22] models are alternative approaches, particularly for disordered or amorphous materials, by estimating the minimum thermal conductivity based on random-walk theory. As these models capture the diffusive heat transport limit, they cannot predict the correct temperature behaviour of the thermal transport over the whole temperature range when thermal transport via phonons is important.[23] Machine learning (ML) approaches have gained popularity due to their ability to predict $\kappa_L$ for certain compounds at a reasonable computational cost.[24–26] In general, ML uses available datasets, either computational (mainly *ab initio*) or experimental data. However, the accuracy of ML depends on the quality of the data used to train the models, which can also limit their application. Accurate phonon properties require well-converged quantum chemical calculations. Training models from scratch for each composition makes the high-throughput use of such models unfeasible. However, cheaper, pre-trained alternatives, so-called foundation machine learned interatomic potentials (MLIP), have recently emerged.

Motivated by the interesting ionic and thermal transport properties and the open questions concerning thermal conductivity models, in this work, we go beyond the simple investigation of the three systems $Ag_8TS_6$ (T = Si, Ge, Sn) with ab initio and experimental approaches and attempt to validate a comparably low-cost, fully ab initio model for thermal conductivity that might be suitable for high-throughput investigations. We build on the recently introduced two-channel model introduced by Xia[27] that incorporates both phonon and diffuson contributions by harmonic phonons and assumes that each phonon lifetime is half of its vibration period. The Xia model simplifies the full lattice-dynamics approach introduced by Simoncelli et al.[28] and is also connected to the analytical two-channel model by Bernges et al.[14] which can be used to fit experimental data. One drawback of the model by Xia is that it has a simplified estimation of phonon lifetimes. To improve the description of the phonon-phonon scattering of each phonon mode, we combine Bjerg's[29] model for computing phonon lifetimes ($\tau$) based on the ideas of Slack, and Xia's two-channel model. This offers a more versatile framework for predicting and analyzing heat conduction, particularly in materials with significant Grüneisen parameters or large unit cells. We also compare it with lattice dynamics calculations based on a foundational machine-learned interatomic potential. Foundational models already offer a cost-effective alternative to ab initio calculations of harmonic phonons.[30] It has also recently been shown that they can reproduce thermal conductivity acceptably for simple binary systems and are compatible within a factor of 2 with ab initio results. With the help of a few additional data points, they can sometimes be fine-tuned for an accurate reproduction of thermal conductivity.[31] Once accurately trained, MLIPs can be used to predict and investigate the thermal conductivity of lattices without performing expensive full *ab initio* calculations of phonon lifetimes based on the relaxation time approach, as implemented in phono3py,[32,33] or without using the costly ab initio Green-Kubo approach.[34–36]

This study fulfils three purposes. First, we demonstrate the clear connection between the bonding properties and thermal conductivity for all three compounds. Essentially, analysing the bonding properties is sufficient to conclude that all three compounds have similar thermal conductivity. Furthermore, we demonstrate that cheap *ab initio* methods, partly combined with machine learning, can analyse and predict lattice thermal conductivity with high accuracy, potentially enabling high-throughput predictions in the future. Lastly, we investigate the relationship between thermal and ionic conductivity.

Therefore, we begin with a detailed quantum-chemical analysis of the bonding in $Ag_8TS_6$ (T = Si, Ge, Sn). Next, we analyse the harmonic phonon properties, including sound velocity ($v$), the Debye temperature, and the volume-dependent Grüneisen parameters ($\gamma$), using both experimental and theoretical methods, and connect these to the bonding analysis. Furthermore, we use the two aforementioned approaches (Grüneisen-based lifetime estimation and foundation model) to predict thermal conductivity and reproduce experimental results. Based on an accurate model of experimental thermal conductivity results and ionic conductivity measurements, we demonstrate that ionic and thermal conductivity are independent in $Ag_8TS_6$ (T = Si, Ge, Sn). By doing so, we also demonstrate the importance of the diffuson channel for these compounds. By integrating bonding analysis, phonon property prediction, and advanced modelling techniques, we aim to establish a robust framework for predicting thermal conductivity inorganic materials, which has implications for the high-throughput discovery of materials.

## Results and Discussion

### Structural description and X-ray diffraction

Single crystal X-ray diffraction reported by Slade et al.[16] revealed that the canfieldite $Ag_8SnS_6$ presents two phase transitions: a low-temperature transition from the orthorhombic phase (space group $Pmn2_1$) (**Figure 1 a, b**) to another orthorhombic phase (space group $Pna2_1$) at 120 K (**Figure 1 c, d**), and a high-temperature transition from orthorhombic $Pna2_1$ to the cubic phase with space group $F\bar{4}3m$ around 460 K. However, in the case of the powder sample, no change in diffraction patterns was observed below 120 K in their study.[16] For the related compounds $Ag_8GeS_6$ and $Ag_8SiS_6$ (**Figure S1**), only the orthorhombic $Pna2_1$ structure has been reported at room temperature.[9,37,38] A detailed report of the coordination environments of all the argyrodites studied here is presented in **Section S2** in the SI. Key results will be discussed as part of the bonding analysis.

In this study, we synthesized $Ag_8TS_6$ (T = Si, Ge, Sn) via a solid-state synthesis approach, and Rietveld refinements of their powder X-ray diffraction patterns at room temperature confirm the formation of single-phase materials (**Figure S2**). Subsequently, temperature-dependent powder X-ray diffraction studies were conducted to investigate the presence of any phase transitions within the temperature range of 100 K to 400 K (**Figure S3**). The Rietveld refinements of all diffraction patterns indicate that the orthorhombic phase, having space group $Pna2_1$, remains stable for both $Ag_8GeS_6$ and $Ag_8SiS_6$ throughout the examined temperature range. Similarly, for $Ag_8SnS_6$, no clear change in the diffraction patterns was observed around 120 K. This may be because the structural variations are too small to detect the low-temperature structural change reported in the literature.[16] Nevertheless, the refined unit cell volume of $Ag_8SnS_6$ below 120 K deviates slightly from a linear increase, suggesting that some structural change may occur at low temperature (**Figure S4**). In contrast, a linear increase in unit cell volume for the other compositions was observed with increasing temperature (**Figure S4**). We shortly note here that we predicted a potential additional phase of $Ag_8SnS_6$, so far not known from experimental work, via ab initio calculations (see **Section S7** in the SI) that was also not found within the experimental investigation. This prediction might be an artefact of the density functional theory (DFT)-based methodology.

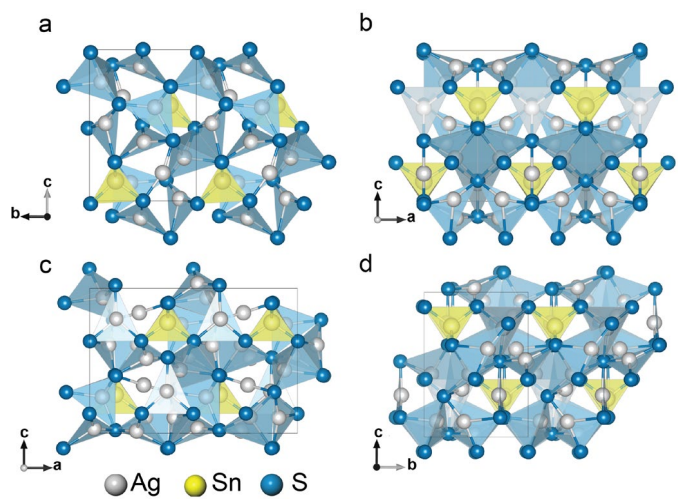

**Figure 1**. Crystal structure of the $Ag_8SnS_6$ compound in the a), b) orthorhombic $Pmn2_1$ space group at 120K (low-temperature) and c), d) orthorhombic $Pna2_1$ space group at room-temperature. These crystal structures form the basis for the computational analysis discussed in the following sections. The crystal structures for the $Ag_8SiS_6$ and $Ag_8GeS_6$ are presented in **Figure S1** of the Supplementary Information.

**Bonding analysis**

Based on the composition alone, we might naively expect $Ag^+$, $Si^{4+}$, $Ge^{4+}$, $Sn^{4+}$, and $S^{2-}$ ions. A closer inspection of the structure, however, was already done by Krebs et al.[37], and suggests the following ionic formula $Ag_8(SiS_4)S_2$ indicating $SiS_4^{4-}$ polyanions isovalent to $SiO_4^{4-}$. From previous bonding analysis results, we also expect very weakly bonded Ag atoms and highly covalent bonds from Si, Ge, and Sn to S. [13,15,16,38,39]

Typically, the bonding situation in a material is used to estimate the sound velocities and to obtain information about the anharmonic nature of the heat transport.[40] For example, bond heterogeneity is typically made responsible for high phonon-phonon scattering rates, and, therefore, low thermal conductivities.[15] Specifically, in the case of $Ag_8SnS_6$, the rattler-like behavior of $Ag^+$ is expected due to the very weak Ag—S bonds.[15] Therefore, we provide a detailed analysis of the bonding situation in all three $Ag_8TS_6$ compounds (T = Si, Ge, Sn) by means of Crystal Orbital Hamilton Populations[41] and Crystal Orbital Bond Orders[42]. Beyond this, we also provide an analysis of metal-metal and multi-center interactions in these compounds, as they might be connected to the overall weak Ag—S bonds.[15]

The bonding situation in all three $Ag_8TS_6$ (T = Si, Ge, Sn) compounds is very similar. The T—S bonds are by far stronger and more covalent than the Ag—S bonds, indicated by both the ICOHP and ICOBI values (**Figure 2a, b**). They also confirm the polyanionic nature of the $TS_4^{4-}$ units, i.e., strong covalent bonds between T and S. The very covalent Sn—S bonds in $Ag_8SnS_6$ show an average ICOHP value of –4.58 eV and an average ICOBI value of 0.84 (close to the ideal ICOBI of 1 of a single bond). In contrast, the Ag—S interactions are much weaker, and the ICOHPs range from –0.66 to –1.61 eV (ICOBIs from 0.12 to 0.34). In the case of the COHPs, occupied antibonding states below the Fermi energy level weaken the Ag—S bonding interactions (see **Figure 2c**). Specifically, Ag (4d) and S (3p) interactions contribute to the antibonding states. Likely due to weak Ag—S bonds, a large number of different, very distorted $Ag^+$ environments exist. We found linear, trigonal planar, trigonal non-planar, and tetrahedral coordination environments for Ag in $Ag_8SnS_6$ (see **Figure S8** in SI). This again confirms the expectation of the mobile nature of the $Ag^+$ ions based on the bonding situation. In contrast, Sn only shows a nearly perfect tetrahedral environment. Besides cation-anion bonds, we also found Ag—Ag interactions in $Ag_8SnS_6$, with ICOHPs ranging from –0.24 to –0.32 eV (ICOBIs from 0.05 to 0.07), likely leading to additional distortions of the Ag environments and weakening of the Ag—S bonds. The exact bond strengths and environments for all $Ag_8TS_6$ (T = Si, Ge, Sn) can be found in **Figures S6-S9 and Table S8-S11** in the SI.

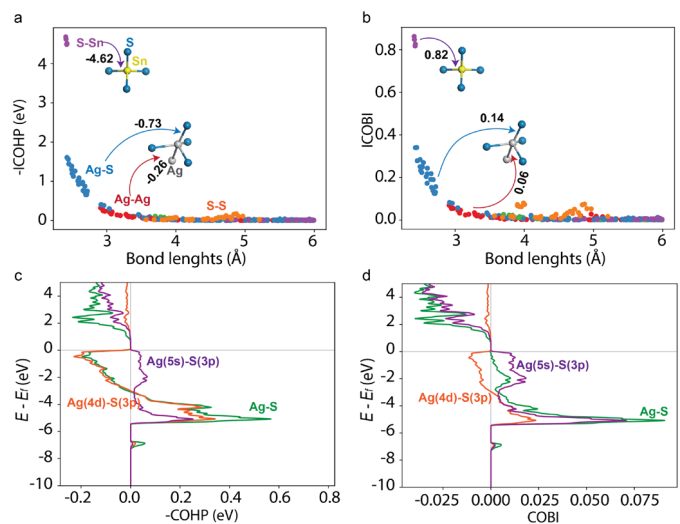

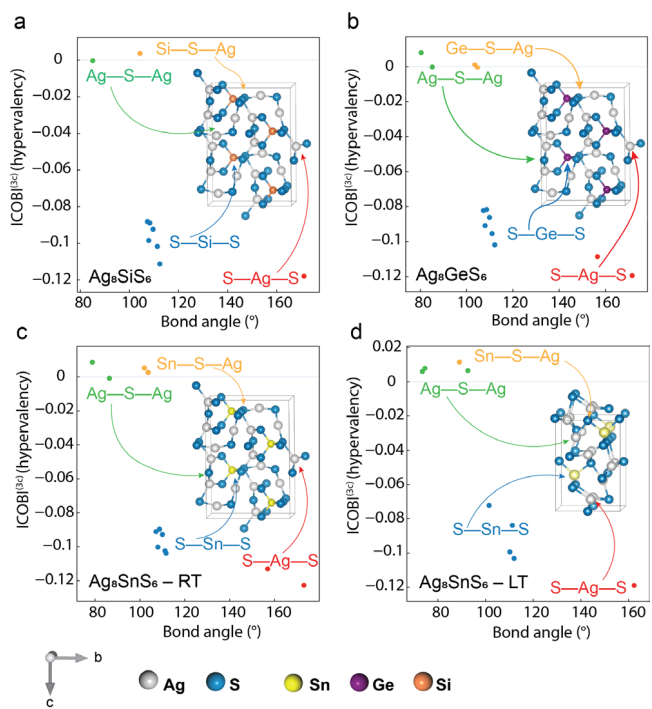

**Figure 2.** a) and b) show the distribution of ICOHP and ICOBI for the RT $Ag_8SnS_6$ structure, respectively. c) and d) depict the weakly bonded Ag—S COHP and COBI interactions at distinct Ag sites. Bonding interactions mainly involve Ag(5s/4d) and S(3p) orbitals, while the antibonding interactions below the Fermi level are dominated by Ag(4d) and S(3p) orbitals.

Plotting all two-center ICOBI$^{(2c)}$ against each compound's bond length unveils an interesting pattern. The ICOBI vs. bond length curve would fall monotonously in a regular compound without a unique bonding situation. Instead, we see unusually strong outliers for bond lengths beyond 3.5 Å (**Figure 2b**). As previously shown in the literature,[43] these outliers indicate potential (hypervalent) multi-center interactions (a detailed discussion can be found in **Section S2** in the SI). This is further investigated, i.e., the three-center (3c) bonds of consecutive atoms with stronger two-center ICOBI (ICOBI$^{(2c)}$ ≥ 0.25) have been taken into account. To get a better overview of the exact bonding situation, all atoms with significant three-center bonds formed by two consecutive bonds with significant ICOBI$^{(2c)}$ as selected above are shown in the structure inset of **Figure 3**, revealing a bonding network that poses a rather untypical bonding situation. **Figure S10** in the SI shows a more detailed picture of the three-center interactions. The ICOBI$^{(3c)}$ is plotted against the bond angle of S with Ag, Sn, Ge, or Si.

**Figure 3.** Three-center ICOBI vs. bond angle plot of $Ag_8TS_6$ (T = Sn, Ge, Si).

As the three-center ICOBI$^{(3c)}$ corresponds to the hypervalency of the bonding electrons, negative values indicate electron-rich, and positive values correspond to electron-poor interactions.[42] It is noticeable that the interactions roughly split up into two categories: weak electron-poor and comparably strong electron-rich bonds. The electron-poor bonds are close to ICOBI$^{(3c)}$ = 0 and consist of Ag—S—Ag and Ag—S—T bonds, while the stronger electron-rich interactions exhibit ICOBIs$^{(3c)}$ between roughly –0.08 and –0.12. Here, it is striking that the S—T—S tetrahedral bonds show comparably strong bonds$^{(3c)}$, further contributing to their covalent character, even though the linear S—Ag—S bonds are favored. Hypervalency has been found and discussed in many polyanions and might therefore not be surprising here. However, it might be unexpected for a $TS_4^{4-}$ polyanionic unit when assuming single bonds and an oxidation state of –2 for S, as the octet rule would be perfectly fulfilled. We have to keep in mind that, although quantitatively, the S—Ag—S and S—T—S bonds seem to have the same strength, in the context of their chemical environment, the bonds differ qualitatively. Compared to extended bonds like S—Ag—S, an ICOBI$^{(3c)}$ of around –0.1 in a local structure element like a tetrahedron can be seen as weak.[43] There is almost no significant difference in the three-center bonds of $Ag_8SnS_6$, $Ag_8GeS_6$, and $Ag_8SiS_6$, except that the Si-analogue shows fewer relevant S—Ag—S multi-center interactions than the other compounds, and therefore is a less dense bond network. A more detailed discussion and comparison to GeTe[44] can be found in the **Section 2** in the SI . It can be assumed that the weak Ag—S bonds, the Ag—Ag interactions, and the S—Ag—S multi-center interactions are closely related and therefore responsible for the anharmonicity of the compounds.

From the bonding analysis results, we see the overall bonding character stays the same when changing the tetrel species. As is known from simple binary compounds, bond strength and sound velocities are typically correlated.[45] Additionally, bond heterogeneity because of rattler-like atoms typically leads to high phonon-phonon scattering and anharmonicity. These similar results for all three compounds therefore suggest that all materials will present very similar sound velocities and high anharmonic transport behavior. Consequently, they are expected to exhibit similarly low lattice thermal conductivities and comparable features in their phonon band structures.

**(Quasi-)harmonic phonon band structures**

Checking the thermal stability of the thermoelectric materials is essential. Commonly, a lack of imaginary modes in the phonon band structure indicates dynamic stability of the structure. For all the argyrodites $Ag_8TS_6$ ($T$ = Si, Ge, and Sn), the phonon frequencies along high-symmetry directions of their Brillouin zone and phonon density of the states (PDOS) do not exhibit imaginary modes, which means that they are dynamically stable. The low-temperature (LT) canfieldite $Ag_8SnS_6$ phase, using a 30-atom unit cell, has 120 phonon modes in total; while the $Pna2_1$ phases of $Ag_8TS_6$ ($T$= Si, Ge, and Sn), with 60 atoms per unit cell, have 180 modes. The phonon dispersion curve also shows considerable overlap between bands, indicating a possible high anharmonicity and a possible diffuson-dominated thermal transport (**Figure 4** and **Figure S5** ).[12,17,22] Here, also the PDOSs show that the Si/Ge/Sn atoms make a small contribution across the entire region, while S atoms mainly dominate the optical frequencies. The acoustic modes produce a dominant peak in the frequency range of 1.6 and 1.8 THz, which corresponds, due to their quantity and atomic mass, to the $Ag^+$ vibrations. Overall, no significant difference was found for the three compounds $Ag_8TS_6$ ($T$ = Si, Ge, Sn) sharing the same crystal structure type.

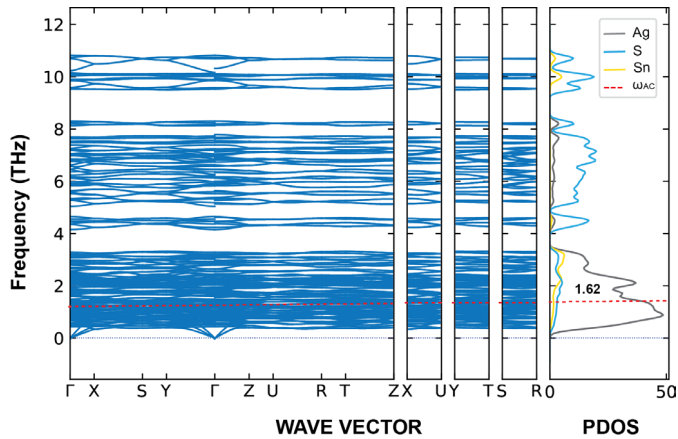

**Figure 4**. Computed phonon band structure along the partial phonon density of states for the room-temperature phase of the $Ag_8SnS_6$ canfieldite. Here, the dotted red line corresponds to the acoustic Debye frequency ($\omega_{AC}$).

**Sound and group velocities**

Sound velocities and Debye temperatures for $Ag_8SnS_6$, $Ag_8GeS_6$, and $Ag_8SiS_6$ were calculated through elastic properties simulations (bulk and shear modulus). In general, related argyrodites with low lattice thermal conductivity, e.g., selenides and tellurides, exhibit mean sound velocities between 1000 and 1500 m/s.[46] The computed mean sound velocities (**Table S14**) show a slight decrease with increasing atomic mass, having a good agreement with the measured mean sound velocities ($v_m^*$). As the bonding analysis above already suggested, we find no significant differences in the speed of sound for the three different compounds. Therefore, the thermal conductivities derived from models based solely on sound velocities and material densities will be almost identical (**Table S15).**

To complement the sound velocity analysis, we also obtained group velocities from harmonic phonon calculations. Again, no significant difference between the argyrodite compounds was observed. Here, the phonon group velocities (**Figure 5**) show higher velocities for low-frequency modes, which are mainly dominated by $Ag^+$ ions vibrations due to their low bonding interaction. Therefore, no substantial distinctions between the compounds from the sound velocities and group velocities can be concluded.

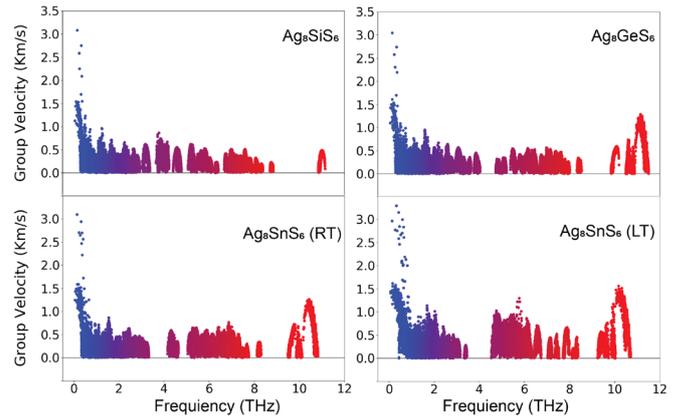

**Figure 4.** Phonon group velocity for $Ag_8TS_6$ ($T$= Si, Ge, and Sn). Notably, higher group velocities are observed at the low-frequency region, which are again mainly dominated by the $Ag^+$ vibration and may influence the low lattice thermal conductivity behaviour.

The Debye temperature and frequency, estimated from both theoretical and experimental results, yield low values as an indication of low lattice thermal conductivity, which is in line with Slack's theory.[18–20] The calculated Debye temperatures and frequencies **(Table S13)** for $Ag_8TS_6$ ($T$= Si, Ge, and Sn) also show a gradual decrease with increasing atomic mass. Nevertheless, the variations in both experimental and theoretical values are minor and do not indicate any significant differences between the three different compounds. Even room-temperature and low-temperature $Ag_8SnS_6$ show a very similar tendency.

**Grüneisen parameter**

In general, the lattice thermal conductivity in a solid depends mainly on the heat capacity, speed of sound, and phonon relaxation time. Materials with low heat capacity, low group velocity, and short phonon lifetime have low lattice thermal conductivity. Both group velocity and phonon lifetime may depend on the bonding situation in the crystal. So far, we have found that $Ag_8SnS_6$, $Ag_8GeS_6$, and $Ag_8SiS_6$ all have weak Ag—S bonds and associated low sound velocities corresponding to $Ag^+$ vibrations. Furthermore, we also expect high anharmonicity of $Ag^+$ vibrations from the bonding analysis.

In order to quantify and evaluate the anharmonicity as a function of the phase and composition, we also compute the variation of the phonon frequencies with respect to the volume change as mode-dependent Grüneisen parameters and derived average quantities ($\gamma$). Given our previous results, we expected larger Grüneisen parameters for all three compounds, but no considerable differences between them. **Figure 6** shows strong anharmonicity represented by a large Grüneisen parameter for the low-energy vibrational modes (highlighted with grey), which are mainly dominated by $Ag^+$ ions. This agrees with the expected mobile/rattler-like nature of the $Ag^+$ ions. The averaged Grüneisen parameter was computed across all modes, showing good agreement with our experimental Grüneisen parameter derived from sound velocity measurements and the one reported previously in the literature ($Ag_8GeS_6$).[13] Despite the change in composition, no significant differences were observed among the experimental average Grüneisen parameters derived from the experimental sound velocities.

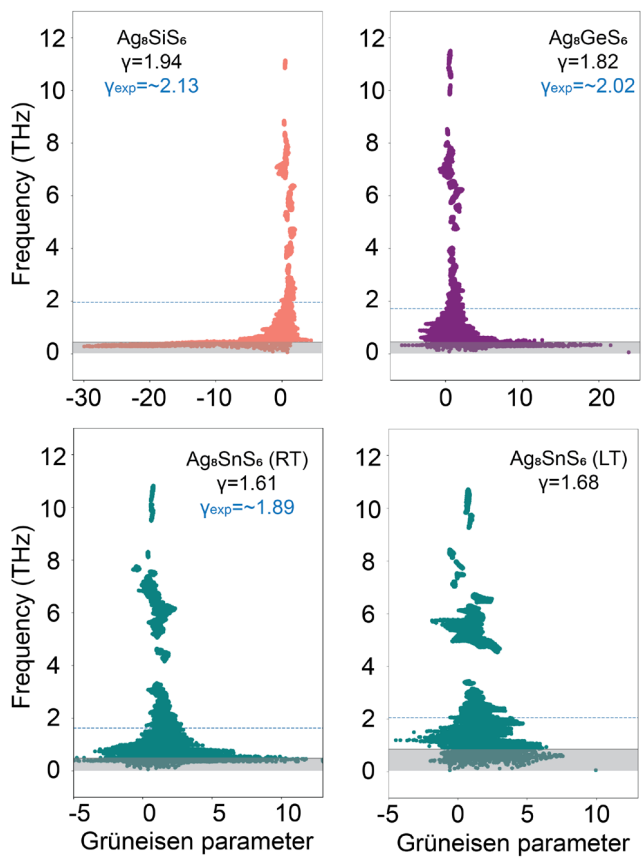

**Figure 6.** Computed mode Grüneisen parameter as a function of frequency for $Ag_8SiS_6$, $Ag_8GeS_6$, and $Ag_8SnS_6$ at room- and low-temperature. Here, we highlight the acoustic modes (grey color) where the anharmonicity is larger. Computed average Grüneisen parameter ($\gamma$) and experimental Grüneisen ($\gamma_{exp}$) are also shown for all investigated structures.

Although we observed comparable experimental-theoretical average Grüneisen values among the compounds, some differences can be observed at lower frequency modes for our theoretical results. For instance, $Ag_8SiS_6$ mostly shows negative Grüneisen parameters for the lower frequencies. The calculation of the average Grüneisen parameter, shown in Figure 6, was performed over all modes. Nevertheless, the average Grüneisen parameters used in the lattice thermal conductivity calculation were calculated with the acoustic modes only, as we expect them to be most important for thermal transport. A comparison of Grüneisen parameter computed over all modes, acoustic modes and up to the Debye frequency are presented in **Figure S17b** in the SI.

**Lattice thermal conductivity**

Various models to predict lattice thermal conductivity have been developed. These range from simple empirical relationships to complex quantum mechanical calculations. These models vary strongly in required computational resources and also in how they model the heat transport in complex solids –either via phonons or diffusons.

Cahill[21] and Agne[22] have developed two alternative models that can be used cost-effectively with ab initio data. When combined with elastic properties obtained from DFT calculations, these models predict minimum lattice thermal conductivity. In both models, the amorphous solid has been used as a model system for the minimum thermal conductivity of crystalline materials; they both rely on random-walk theory, indicating heat transport in amorphous materials via diffusons. Because of this, these models can only be used in the high-temperature limit of crystalline materials.

Slack[18–20], on the other hand, provides a lattice thermal conductivity model based on heat transport via phonons and as a function of temperature. This model emphasizes the role of the acoustic phonon modes in the thermal transport processes. According to Slack, the lattice thermal conductivity is influenced by factors such as the Debye temperature, sound velocity, and the Grüneisen parameter, which accounts for the anharmonicity of the lattice vibrations. The model is particularly useful for estimating the upper limit of thermal conductivity in crystalline materials with strong atomic bonding. In this approach, the lattice thermal conductivity can be computed as:

$$\kappa_{\text{Slack}} = A \frac{\bar{M} \delta n^{1/3} \Theta^3}{\gamma^2 T} \qquad (1)$$

Where $\bar{M}$ is the average atomic mass, $V$ is the volume of the unit cell, $\Theta$ is the acoustic Debye temperature, $T$ is the absolute temperature, $k_B$ and $\hbar$ are the Boltzmann and Planck constants, respectively, and $A$ is the Slack coefficient, which is dependent on the anharmonicity of the structure, represented by the average Grüneisen parameter $\gamma$.

$$A = \frac{2.436 \times 10^{-8}}{1 - \frac{0.514}{\gamma} + \frac{0.228}{\gamma^2}} \qquad (2)$$

The Slack model is valuable as it provides a more thorough temperature-dependent analysis of thermal conductivity, offering insights that other models may not capture, especially in materials where acoustic phonons play a dominant role. While it yields important information about lattice thermal conductivity, the model also has limitations. Many studies point out that lattice thermal conductivity is generally overestimated when compared with experimental data. This discrepancy can be related to the $A$ coefficient. Qin and coworkers[47] address this problem by scaling the $A$ coefficient or by fitting the $A$ parameter.

A recent model for high-throughput screening and analysis of thermal conductivity was introduced by Xia et al.,[27] where the lattice thermal conductivity can be estimated through harmonic phonon calculations. This model provides a complementary perspective to the previous methods since it builds upon the so-called two-channel model (phonon-gas channel and diffuson channel) where the total thermal conductivity $\kappa_l$ is calculated from the sum of the phonon (s=s') and diffuson contributions (s≠s') (Eq. 3).

$$\kappa_l = \sum_{qss'} C_{ss'}(\boldsymbol{q}) v_{ss'}(\boldsymbol{q}) v_{s's}(\boldsymbol{q}) \tau_{ss'}(\boldsymbol{q}) \qquad (3)$$

$C_{ss'}(\boldsymbol{q})$ is a heat capacity matrix element, $v_{ss'}(\boldsymbol{q})$ is a velocity matrix element and $\tau_{ss'}(\boldsymbol{q})$ is a phonon lifetime matrix element for two phonons at the branches $s$ and $s'$ in reciprocal space at $\boldsymbol{q}$. $\tau_{ss'}(\boldsymbol{q})$ can be computed based on $\Gamma_s(\boldsymbol{q})$ – the scattering rate or the inverse of phonon lifetime of the phonon at branch $s$ at point $\boldsymbol{q}$ – and $\omega_s(\boldsymbol{q})$, its frequency:

$$\tau_{ss'}(\boldsymbol{q}) = \frac{2(\Gamma_s(\boldsymbol{q}) + \Gamma_{s'}(\boldsymbol{q}))}{4(\omega_s(\boldsymbol{q}) - \omega_{s'}(\boldsymbol{q}))^2 + (\Gamma_s(\boldsymbol{q}) + \Gamma_{s'}(\boldsymbol{q}))^2} \qquad (4)$$

This two-channel approach, introduced by Simoncelli et al.,[28,48] has been very useful when disordered materials or crystals with large unit cells, such as $Yb_{14}Mn_1Sb_{11}$, have been investigated.[49] Simoncelli's model relies on the ab initio computation of phonon lifetimes, which can be computationally very demanding, making it extremely expensive for a large-scale screening approach. In contrast to this, the model by Xia.[50] purely relies on harmonic phonon calculations, making it significantly more affordable. It additionally assumes that each phonon lifetime $(1/\Gamma_s(\boldsymbol{q}))$ is half of its vibration period.

A comparison of lattice thermal conductivity using the models mentioned above is shown in **Table S15**. All models consistently predict the material's low lattice thermal conductivity, which can be attributed to its weak bonding, low sound velocities, and the high anharmonicity of the low-energy vibrational modes dominated by $Ag^+$ ions. Furthermore, the diffusion-mediated minimum conductivity ($\kappa_{\text{Agne}}^{\min}$) and two-channel ($\kappa_{\text{Xia}}^{\min}$), both indicate that heat conduction is primarily dominated by diffusons, as it was also shown in previous studies including sulfide- and selenide-argyrodites.[12,17]

Although, it is clear that all these models predict low minimal lattice thermal conductivities for $Ag_8SnS_6$, $Ag_8GeS_6$, and $Ag_8SiS_6$, a full *ab initio* model that can provide a detailed insight into the thermal properties is missing. Thus, to incorporate the anharmonicity in the prediction of the lattice thermal conductivity and to reduce the overestimation that the Slack model often shows. We start from the two-channel approach proposed by Xia et al. but go beyond the minimum lattice thermal conductivity approximation. We model the phonon lifetimes $(1/\Gamma_s(\boldsymbol{q}))$ using the method proposed by Bjerg and co-workers, which is based on Slack's approach.[18,29] By incorporating inverse phonon lifetimes through the Grüneisen parameter, we effectively account for phonon-phonon scattering, which constitutes the dominant process limiting the lattice thermal conductivity in these materials. Then, the inverse phonon lifetimes are calculated as follows:

$$\Gamma_s(\boldsymbol{q}) = p(\omega_s(\boldsymbol{q}))^2 \frac{T}{\Theta} e^{-\Theta/3T} \quad (5)$$

$\omega_s(\boldsymbol{q})$ is the phonon frequency, and $p$ is a fitting function that is dependent on the average Grüneisen parameter $\gamma$, which can be determined by:

$$p = \frac{1 - 0.514\gamma^{-1} + 0.228\gamma^{-2}}{0.0948} \frac{\hbar^2 \gamma^2}{k_B \Theta \overline{M} V^{1/3} v} \quad (6)$$

Here $v$ is the speed of sound and is determined from the Debye frequency $\omega_D$, the number of atoms, and the volume of the cell, via the following equation:

$$v = \frac{\omega_D}{\sqrt[3]{6\pi^2 \frac{N}{V}}} \quad (7)$$

Following the proposed model in this study, we computed the two-channel temperature-dependent lattice thermal conductivity, where the diagonal components of the heat flux matrix correspond to the phonon contribution, the off-diagonal components correspond to the diffuson contribution, and the total lattice thermal conductivity is obtained by summing both contributions. With this, in **Figure 7b and Figure S18,** we show the ultra-low total thermal conductivity for the sulfide-argyrodite materials with a very good agreement with the experimental measurements in the high temperature range.

For $Ag_8GeS_6$, where additional low-temperature experimental data are available, deviations are observed between 0 and 50 K compared to other measurements. We attribute this deviation to the presence of point-defect scattering, which can be caused when imperfections, such as atomic-scale substitutions, vacancies, or interstitials, disrupt the periodicity of the crystal lattice. This disruption could create a barrier to phonon propagation, significantly reducing lattice thermal conductivity.[12,14,51,52] Furthermore, the presence of microstructure features, such as grain boundaries, phase segregation, as well as different grain sizes in the experimental samples, can scatter phonons and decrease thermal conductivities and contribute to discrepancies with computational approaches that do not correct for these effects.[14,53]

To estimate the influence of the point defects and the microstructure, we fitted the analytical model described in ref [14] to the experimental data of $Ag_8GeS_6$. This fitting was previously presented and discussed in our earlier work,[13] however, it is also included here to provide a complete comparison between our proposed models. This analytical model also accounts for phonon and diffuson channels and starts from harmonic phonon data as computed by DFT. Below the frequencies of the Ioffe-Regel limit, the Callaway model will be used to describe the heat transport, while above this limit, the model introduced by Agne will be used to describe the diffuson channel. However, estimations of lifetimes within the Callaway model, including effects from point defects and microstructure in the phonon lifetimes, are now fitted to the experimental data. **Figure 7a** shows this fitting, demonstrating that heat transport can be accurately described based on this analytical model. It also indicates that the suppression of the phonon peak is predominantly driven by point-defect scattering and boundary scattering from microstructural characteristics, such as grain size. This analysis can be seen in **Figure 7d**, which highlights the role of point-defect and boundary scattering within the phonon channel. For the Grüneisen-based approach, we observe a divergence in the 0–20 K range. However, the overall features of the temperature-dependent thermal conductivity agree very well with the experiment and especially our analytical model in which effects from point-defect and boundary scattering have been subtracted.

Given the experimental uncertainty, the foundation machine-learned interatomic potential (MACE-MP-03b) reached good results in comparison with experiments as well, as shown in **Figure 7c/d**. To compute the thermal conductivity, we use the full two-channel lattice dynamics approach implemented by Simoncelli et al.[28,48] **Figure 7d** and **Figure S19**, respectively, compare the phonon-channel and total lattice thermal conductivity obtained from the ML model and the Grüneisen parameter-based estimation. The result from the ML model agrees very well with the analytical model over the whole temperature range when point-defect and boundary scattering are subtracted. This again highlights the importance of point defects and boundary scattering for an accurate description of the thermal conductivity. Overall, the ML potential yields result consistent with the Grüneisen model, demonstrating that both approaches reliably capture this system's thermal transport behaviour – even in the low-temperature region.

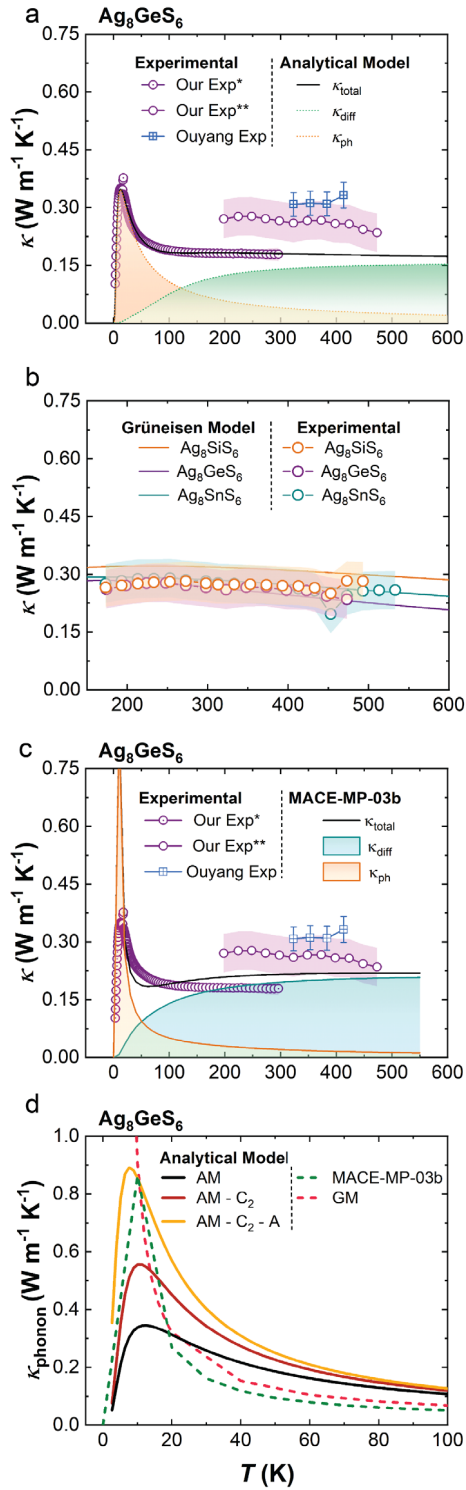

**Figure 7.** a) Fit of the low-temperature measured thermal conductivity data (Experimental**) of $Ag_8GeS_6$. An additional measurement from literature[17] and results from a second method to measure thermal conductivity (this study Experimental*) are also provided in the plot for comparison. The fit is performed with the help of the analytical model as proposed in ref [14] (scattering coefficients are presented in **Table S18** in the SI. b) Comparison of the lattice thermal conductivity following our proposed Grüneisen Model (GM) with experimental measurements for $Ag_8TS_6$ ($T$ = Si, Ge, Sn). c) Two-channel lattice thermal conductivity using the foundation model MACE-MP-03b. d) Contribution of scattering process, phonon-phonon scattering ($C_1$), point-defect ($C_2$), and boundary scattering (A) on the phonon channel as obtained from the analytical model, compared with the Grüneisen model and the foundation model MACE-MP-03b. Although the two proposed approaches show minor differences, they remain consistent with the experimental results within a three-fold standard deviation, showing especially strong agreement for temperatures above 200 K.

Overall, following our proposed models, the results align with the findings of Ouyang and coworkers,[17] confirming that heat transport in the argyrodites ($Ag_8GeS_6$ and $Ag_8SnS_6$ (RT)) is dominated by heat transport in the diffuson-channel. We note that we neglected the influence of four-phonon scattering processes and additional temperature renormalizations of the harmonic phonons that slightly influence the results, in contrast to the simulations by Ouyang and coworkers.[17] All theoretical predictions and the experimental results reveal no significant differences among the three compositions $Ag_8SiS_6$, $Ag_8GeS_6$, and $Ag_8SnS_6$. The Grüneisen-based model, however, results in a slight difference between the thermal conductivity of the $Ag_8SiS_6$ and $Ag_8SnS_6$ compounds, which also corresponds to the differences observed in the computed Grüneisen parameter results.

As the Grüneisen-based model is computationally comparably cheap and a foundation MLIP model even requires less computational cost, they would both be suited for a high-throughput approach for screening thermal conductivity. However, it is currently unclear for which composition spaces foundational ML potentials might fail and how cheap finetuning for complex systems could look like. First finetuning tests with additional ab initio data from rattled supercells with an average displacement of 0.1 Å worsened the description of the phonon channel in our case, while the harmonic phonon results improved. We hope that automated MLIP training and finetuning capabilities will support establishing efficient training and finetuning procedures.[54,55] Despite these challenges, MLIPs are very promising here as MLIPs allows for a full ab initio calculation of the lifetimes and including temperature renormalization effects in the phonons or four-phonon processes comparatively easily. Both the Grüneisen und MLIP approaches could also be combined to spot systematic failures of the foundation model within a high-throughput approach, or the Grüneisen model might be used together with a foundation MLIP. For heat capacity simulations, we have seen previously that even ML models with comparably poor predictions can lead to good heat capacity estimates when constrained enough by a physical model.[56]

As in the previous studies on the argyrodite such as $Ag_8GeSe_6$, $Cu_7PSe_6$, and $Ag_{8-x}Cu_xGeS_6$,[8,12–14] the thermal conductivity can also be modelled without considering changes in ionic conductivity. In those studies, the low thermal conductivity and the high ionic conductivities were shown to be independent of each other, as the ionic conductivities vary drastically within the same temperature range in which the thermal conductivity was modelled purely based on the lattice dynamics simulations. The following section investigates ion transport properties to shed further light on the situation in $Ag_8TS_6$ ($T$ = Si, Ge, Sn).

**Ionic conductivity**

To evaluate the ionic conductivity and its temperature dependence in the argyrodite compounds $Ag_8TS_6$ ($T$ = Si, Ge, Sn), electrochemical impedance spectroscopy measurements were performed over the temperature range 233-303 K. These $Ag_8TS_6$ ($T$ = Si, Ge, Sn) argyrodites are known as mixed ionic-electronic conductors.[57] Therefore, to assess the electronic conductivity, electronic direct current polarization experiments were first carried out. The results indicate that the electronic conductivity of all three argyrodites lies in the range of $0.141 \times 10^{-4}$ mS/cm to 0.0175 mS/cm, confirming minimal electronic contribution (**Figure S14**). To ensure accurate determination of ionic conductivity, an electron blocking, ion conducting electrode ($RbAg_4I_5$)[12,13] was used to suppress electronic interference (details are provided in Section S4 in the SI). Nyquist plots at 233 K for all three compounds are presented in **Figure 8a**. The obtained ionic conductivities are $0.081 \pm 0.007$, $0.065 \pm 0.005$, and $0.075 \pm 0.008$ mS/cm for $Ag_8SnS_6$, $Ag_8GeS_6$, and $Ag_8SiS_6$ respectively at 298 K (errors represent standard deviations from triplicates). The activation energies for ion transport, calculated from Arrhenius plots (shown in the Supporting Information **Figure S16**), are nearly identical for all three argyrodites: $0.29 \pm 0.02$ eV. These results suggest that isovalent substitution at the $T$-site (Si, Ge, Sn) does not significantly affect ionic conductivity, a trend consistent with the behavior observed in the thermal transport properties. Literature study suggests that the relatively low obtained ionic conductivity in these $Ag^+$ conducting argyrodites may be attributed to their crystal structures, where all $Ag^+$ positions are nearly fully occupied.[13,57] The variation of ionic and thermal conductivity with increasing temperature, measured within the same temperature range, is illustrated in **Figure 8b**. Across all compositions, the ionic conductivity increases from an average of ~0.003 mS/cm at 233 K to ~0.1 mS/cm at 303 K, indicating an enhancement of more than one order of magnitude with rising temperature. In contrast, the thermal conductivity remains nearly constant over the same temperature range, varying only marginally between 0.27 W/mK and 0.28 W/mK. This observation suggests that ion transport has no direct influence on the observed low thermal conductivity in these materials, corroborating the previously reported findings for Ag and Cu-based selenide and sulfur argyrodites.[12,8,13]

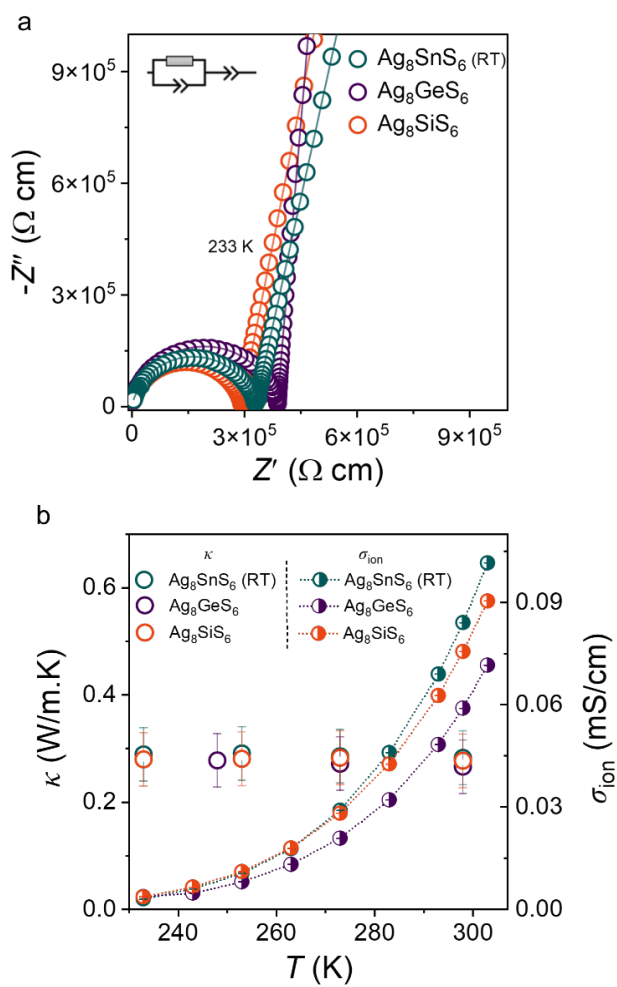

**Figure 8.** a) Normalized Nyquist plots of $Ag_8TS_6$ ($T$ = Si, Ge, Sn) recorded at 233 K, showing higher resistance values, which indicate lower ionic conductivity ($\sigma_{ion}$). b) The variation of thermal ($\kappa$) and ionic conductivity ($\sigma_{ion}$) with temperature, exhibiting no direct correlation between the two.

## Conclusions

Our results demonstrate a strong relationship between chemical bonding and lattice thermal conductivity in Ag-based sulfide argyrodites. The similar bonding strengths in all compounds lead to very similar sound velocities, while the weakly bonded Ag$^+$ atoms result in high anharmonicity of vibrations, associated with high Grüneisen parameters. This weakness likely originates from occupied antibonding states in Ag—S bonds, Ag—Ag bonds, and the multi-center interactions. This further supports that the bonding situation might be predictive of a compound's overall thermal conductivity.

By applying both the Grüneisen-based and the MLIP-based model, we achieve good agreement with the experimental thermal conductivity data, especially in the medium/high temperature range. To capture the characteristic low-temperature peak, it is essential to include point-defect scattering, which effectively suppresses the phonon peak. In addition, microstructural features, most notably grain boundaries, introduce further boundary scattering, with grain size emerging as a key design parameter for tailoring thermal transport. Both effects are shown based on a fit of experimental data with an analytical model.

Overall, these results again demonstrate that accurately modelling heat transport in structurally complex materials over a large temperature range requires capturing the combined influence of bonding-driven anharmonicity, sound velocity, point-defect scattering, and microstructural effects. Furthermore, we identify two approaches that might be suitable for comparably cheap high-throughput screening of lattice thermal conductivity over wide temperature ranges. However, further verification is needed.

## Experimental and Theoretical Work
### Methodology
**Atomistic Simulations**

Electronic-structure computations were performed using Density Functional Theory (DFT) as implemented in the Vienna Ab initio Simulation Package (VASP) [58–60]. The exchange-correlation functional was treated in the semi-local approximation of Perdew, Burke, and Ernzerhof (PBE) with generalized gradient approximation (GGA)[61,62]. The plane wave cut-off was set to 520 eV. To sample the Brillouin zone, we employed Γ-centre grid with a maximum separation of 0.12 Å$^{-1}$, which corresponds to a 7×7×5 and 3×7×5 k-points mesh for the orthorhombic (Pmn2$_1$) and orthorhombic (Pna2$_1$) structures, respectively. Structure optimizations were carried out in terms of volume, cell shape, and ionic positions.

The vibrational properties were computed using the supercell approach with the finite displacement method implemented in phonopy with displacements of 0.01 Å [63,64]. To obtain the dynamical matrix D(q), we used a supercell model of (3×3×2) and (1×3×2) for the LT and RT structures, respectively. The supercell calculations for the LT structure were performed at the Γ-point, while for the RT structure a 3×2×2 Γ-centred k-point grid was needed. In order to correct the dipole interaction, we also employed non-analytical term correction using Born charges as computed with VASP.[65,66]

To compute volume-dependent thermal properties, we employed the Quasi-Harmonic Approximation (QHA)[67], implemented in phonopy.[32,68] To do so, we applied the harmonic approximation at expanded and contracted volumes. We start with the fully optimized structure at the ground state (V$_0$), and then we compute the constant volume energy of 13 different volumes from 0.943 × V$_0$ to 1.063 × V$_0$ in steps of 0.013 × V$_0$. The lattice parameters and atom positions were optimized by minimizing the electronic energy (ISIF=4).[69] Additionally, to compute the anharmonicity of the structures, we compute the Grüneisen parameter. Here, two additional structural optimizations were performed at constant volume, 1% × V$_0$, and −1% × V$_0$.

To obtain Cahill's minimum thermal conductivity, we performed elastic constant calculation using an automated workflow implemented in atomate2, where elastic tensors are computed from stress-strain relationships.[70–73]

To compute the lattice thermal conductivity based on the foundation ML potential (here MACE-MP-03b model[74]) together with the two-channel model introduced by Simoncelli et al.,[28,48] we solved the Wigner transport equation model as implemented in phono3py.[32,68] For this purpose, the third-order force constants were obtained with a supercell of 1×2×2, and the reciprocal space was sampled with a 6×14×10 mesh. Due to very demanding memory requirements, we only used the relaxation time approximation.

To get chemical insight into these compounds, bonding analysis was performed. To do so, we used our recently developed automatic bonding analysis workflow.[75] The fully optimized structure for phonon computations is used as the input structure to start this workflow. The workflow then performs the bonding analysis with the LOBSTER[76–79] program by adding all necessary computational steps to the pipeline. This pipeline consists of a static DFT computation using the GGA functional parameterized by PBE[61,62] within the PAW framework.[80,81] A grid density of 6000 k-points per reciprocal atom is set for the DFT run. The electronic structure's convergence criterion and the plane-wave energy cutoff are set to 10$^{-6}$ and 520 eV, respectively. The number of grid points (NEDOS) on which the density of states is evaluated is set to 10000. The Brillouin zone is integrated using the tetrahedron method with Blöchl[82] correction (i.e., ISMEAR=-5). In all DFT computations, spin polarization is switched on, even though this is not required for these compounds. The workflow also performs LOBSTER computations with the available basis for projecting the wavefunctions. Here, we report the results on the minimal basis.

For bonding analysis runs via LOBSTER, COHPs and COBIs are computed for the entire energy range of VASP static runs, and the COHP/COBI energy interval step is set to 10000 points (equal to NEDOS set in the VASP static run). The increased number of points assigned for the COHP/COBI computation poses a very good estimate of bonding and anti-bonding contribution in bonds during post-processing the results via LobsterPy. Three-center interactions to calculate three-center COBI and ICOBI were chosen according to stronger two-center ICOBIs (cutoff ICOBI$^{(2)}$ = 0.2) of three consecutive atoms and automatically analyzed using a new implementation by one of the current authors in pymatgen (as of v2023.10.11).[83] Other multi-center bonds have been checked as well, but did not yield significant values (cutoff ICOBI(n) = ±0.05).

**Solid-state synthesis of Ag$_8$TS$_6$ (T = Si, Ge, Sn)**

The synthesis of Ag$_8$TS$_6$ (T = Si, Ge, Sn) utilized reactants including silver powder (99.9%, sigma aldrich), silicon (99.999%, Thermo Scientific)), germanium (99.999%, sigma aldrich), tin (99.85%, Thermo Scientific), and sulfur powder (99.98%, sigma aldrich). A high-temperature solid-state synthesis method was conducted under static vacuum conditions to produce polycrystalline samples of Ag$_8$TS$_6$. Initially, stoichiometric amounts of the reactants were weighed inside an argon-filled glovebox and placed into carbon-coated quartz ampoules, which had been pre-dried at 1073 K for 2 hours under dynamic vacuum. These ampoules were then sealed under vacuum and heated in a horizontal tube furnace. The heating process involved ramping the temperature to 523 K at a rate of 50 K per hour, followed by a 24-hour hold. Subsequently, the temperature was increased to 1250 K at the same rate, maintained for 60 hours, and then cooled down to room temperature.

**X-ray diffraction**

X-ray diffraction patterns of Ag$_8$TS$_6$ (T = Si, Ge, Sn) were collected using a STOE STADIP diffractometer. The setup utilized Mo Kα1 radiation ($\lambda$ = 0.7093 Å) equipped with curved Ge (111) monochromator and a Mythen2 1K detector. Measurements were performed in the Debye-Scherrer geometry over a 2$\vartheta$ range from 4° to 44°, at a scan rate of 1° per minute. The temperature range during these measurements was between 100 K and 400 K, maintained using a Cryostream 1000 cooler from Oxford Cryosystems Ltd. for low-temperature conditions (<298 K). Samples were prepared in borosilicate glass capillaries with a 0.5 mm diameter, and they were equilibrated for 20 minutes at each temperature step prior to the measurement. Details of structural phase analysis and Rietveld refinements are discussed in **Section S1** in the SI.

**Ultrasonic speed of sound measurement**

An Olympus Epoch 600 with 5 MHz transducers was employed to measure speed of sound using the pulse-echo method. Variations in signal measurements and the geometrical factors (such as density and thickness) were accounted for to determine the uncertainty of the speed of sound measurement.

**Thermal transport properties measurement**

A Netzsch LFA-467 instrument was used to measure thermal diffusivity of all three compositions, using 10 mm diameter, disc-shaped samples with a bulk density of approximately more than 95% of the theoretical density. Measurements were conducted over a temperature range of 173 K to 500 K. An MCT detector with a ZnS furnace window was used for the measurements below room temperature; while for measurements from room temperature to high-temperature, an InSb detector with a sapphire furnace window was employed. The detection time and signal amplification were optimized automatically for each measurement. At every temperature point, three measurements were taken, with five measurements conducted at 173 K to ensure accuracy. The detector signal was analyzed using an improved Cape-Lehman model.[84–86] All samples were spray-coated with graphite to enhance the infrared light absorption and emission during the laser-flash measurements. The equations used to calculate the thermal conductivity from the measured thermal diffusivity are provided in **Section S3** in the SI. Additionally, low-temperature thermal conductivity measurement for Ag$_8$GeS$_6$ was performed using a Physical Property Measurement System (PPMS) with the TTO option, under high vacuum (<10$^{-4}$ Torr) and with a temperature gradient of approximately 3% between the hot and cold sides. A disc-shaped sample (4 mm× 2 mm) was used for the measurement.

**Direct current (DC) polarization measurements**

DC polarization measurements were performed using a press cell with a 10 mm inner diameter. The samples were filled into the press cell, and stainless-steel stamps were used as ion blocking electrode on both sides. The cells were then closed and subjected to uniaxial pressing at 3 tons for 3 minutes. A VMP-300 potentiostat (Biologic) was used to carry out DC polarization, applying a voltage ranging from 5 mV to 50 mV in 5 mV steps. To ensure equilibrium at each step, the applied voltage was kept constant for 2 hours before proceeding to the next step.

**Electrochemical impedance spectroscopy**

Electrochemical impedance spectroscopy measurements were carried out using a cell set up comprising two stainless steel stamps that served both as current collectors and as a means to press the sample during the measurements. The samples were placed in an insulating PEEK housing with an inner diameter of 10 mm. First, the argyrodite materials were loaded in the PEEK housing and pressed under 3 tons of uniaxial pressure for 3 minutes. Secondly, the cells were opened in a glovebox, and a thin layer of RbAg$_4$I$_5$ was pressed onto both sides of the sample, followed by an additional 5 minutes of compression using a manual screw press. Finally, AC impedance spectroscopy was performed over the temperature range of 233 K -

303 K using an SP300 impedance analyzer (Biologic). The measurements employed an excitation amplitude of 10 mV and covered a frequency range of 5 MHz to 1 Hz. The analysis of impedance results is shown in **Section S5** in the Supporting Information.

## Author contributions

Conceptualization: JB, JG; Data Curation: JB, AN, CE, AG, JG; Formal Analysis: JB, JG, AN, CE, AG, WZ; Funding Acquisition: WZ, JG; Investigation: JB, AN, CE, AG, WZ; Methodology: JB, JG, AN, CE, KU, AG, WZ; Project Administration: WZ, JG; Resources: WZ, JG; Software: JB, JG, AN, CE, KU, AG; Supervision: , WZ, JG; Validation: JB, JG; Visualization: JB, AN, CE, AG; Writing – Original Draft: JB, JG, AG, WZ with contributions from all authors; Writing – Review & Editing: All authors

## Data availability

The raw computational data required to reproduce the results of this study are available through our ZENODO repositories https://doi.org/10.5281/zenodo.17397457 and https://doi.org/10.5281/zenodo.17399976. All post-processing and plotting scripts used to generate the figures presented in this work are openly accessible on our GitHub repository https://github.com/DigiMatChem/paper-grueneisen-model-for-argyrodites/ (https://doi.org/10.5281/zenodo.17437320). Detailed instructions for reproducing each dataset and figure are provided in the accompanying README file.

## Acknowledgements

The research was supported by ERC Grant MultiBonds (grant agreement no. 101161771; funded by the European Union. Views and opinions expressed are however those of the author(s) only and do not necessarily reflect those of the European Union or the European Research Council Executive Agency. Neither the European Union nor the granting authority can be held responsible for them.) The research was additionally supported by the Deutsche Forschungsgemeinschaft (DFG) under grant number ZE 1010/15-1 and project number 459785385. Furthermore, we would like to acknowledge the Gauss Centre for Supercomputing e.V. (www.gausscentre.eu) for funding this project by providing generous computing time on the GCS Supercomputer SuperMUC-NG at Leibniz Supercomputing Centre (www.lrz.de) (Project No. pn73da).

## Notes and references

Supplementary information: Thermal Transport in Ag$_8$$T$S$_6$ ($T$= Si, Ge, Sn) Argyrodites: An Integrated Experimental, Quantum-Chemical, and Computational Modelling Study


*Joana Bustamante,$^a$ Anupama Ghata,$^b$ Aakash A. Naik, $^{a,c}$ Christina Ertural,$^a$ Katharina Ueltzen,$^{a,c}$ Wolfgang G. Zeier,$^{b,d}$ and Janine George *$^{a,c}$*

$^a$Federal Institute for Materials, Research and Testing, Department Materials Chemistry, Unter den Eichen 87, 12205 Berlin, Germany.

$^b$University of Münster, Institute of Inorganic and Analytical Chemistry, Corrensstr. 28/30, D-48149 Münster, Germany.

$^c$Friedrich Schiller University Jena, Institute of Condensed Matter Theory and Solid-State Optics, Max-Wien-Platz 1, 07743 Jena, Germany.

$^d$Institute of Energy Materials and Devices (IMD), IMD-4: Helmholtz-Institut Muenster, Forschungszentrum Jülich, 48149 Münster, Germany.

E-mail: janine.george@bam.de




# Content





**Section S1: Phase analysis and Rietveld refinements of Ag$_8T$S$_6$ ($T$ = Si, Ge, Sn)**

Rietveld refinements were performed at all temperature steps for all three compositions of Ag$_8T$S$_6$ ($T$ = Si, Ge, Sn) using the Topas-Academic V7 software package.[1] A Chebyshev polynomial function was used to model the background, whereas the peak shapes were described using a modified Thompson-Cox-Hastings pseudo-Voigt function.[2] At first, refinements were carried out for background coefficients, sample displacement, lattice parameters, and peak shapes. After that, the fractional atomic coordinates and the isotropic thermal displacement parameters of atoms were refined. The refinements proceeded sequentially, starting with sulfur (S$^{2-}$), moving to $T^{4+}$ ($T$ = Si, Ge, Sn), and finally silver (Ag$^+$). To reduce the number of free variables in the refinements, the isotropic thermal displacement parameters for all silver (Ag$^+$) atoms were constrained to be equal across all three compositions. Similarly, the isotropic thermal displacement parameter for $T^{4+}$ ($T$ = Si, Ge, Sn) atoms and all sulfur atoms (S$^{2-}$) were also considered to be equal for better refinement quality.

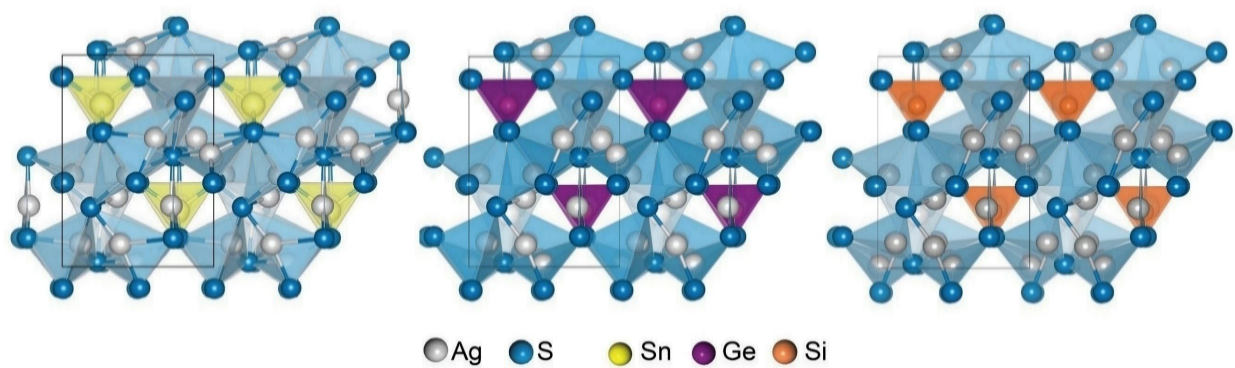

*Figure S1*. Lateral view of Ag$_8$SiS$_6$, Ag$_8$GeS$_6$, and Ag$_8$SnS$_6$ argyrodites in the orthorhombic $Pna2_1$ phase at room temperature.

The results of Rietveld refinements of Ag$_8T$S$_6$ ($T$ = Si, Ge, Sn) at 298 K are presented in **Tables S1-S3**, and **Figure S2**, while the X-ray diffraction patterns at all measured temperatures are shown in **Figure S3**. The refined lattice parameters and unit cell volume for all compositions are also tabulated (**Table S4-S6**). From the slope of the unit cell volume vs. temperature plots (**Figure S4**), the thermal volume expansion coefficients were calculated for each composition. The $R_{wp}$ and the goodness-of-fit (GoF) value indicate the refinement quality.



*Table S1. Structural parameters of Ag$_8$SiS$_6$ at 298 K as obtained by Rietveld refinements and utilizing laboratory X-ray diffraction (Mo Kα radiation).*

| Structural information of Ag$_8$SiS$_6$ from X-ray diffraction data at 298 K |||||||
|---|---|---|---|---|---|---|
| Space group: $Pna2_1$; λ (Mo Kα) = 0.7093Å; |||||||
| Lattice parameters: $a$ = 15.058 (1) Å, $b$ = 7.4355(6) Å, $c$ = 10.5415(9) Å, |||||||
| $R_{wp}$ = 7.81%; GoF = 2.11 |||||||
| **Atom** | **Wyckoff site** | $x$ | $y$ | $z$ | Occ. | $B_{eq}$ / Å$^2$ |
| Ag1 | 4$a$ | 0.01681(5) | 0.011(1) | 0.016(1) | 1 | 3.53(9) |
| Ag2 | 4$a$ | 0.0640(4) | 0.2287(8) | 0.255(1) | 1 | 3.53(9) |
| Ag3 | 4$a$ | 0.1244(4) | 0.2236(8) | 0.787(6) | 1 | 3.53(9) |
| Ag4 | 4$a$ | 0.2206(5) | -0.006(1) | -0.006(1) | 1 | 3.53(9) |
| Ag5 | 4$a$ | 0.2607(4) | 0.114(1) | 0.305(2) | 1 | 3.53(9) |
| Ag6 | 4$a$ | 0.2690(5) | 0.380(1) | 0.091(1) | 1 | 3.53(9) |
| Ag7 | 4$a$ | 0.4194(5) | 0.098(1) | 0.107(1) | 1 | 3.53(9) |
| Ag8 | 4$a$ | 0.4327(6) | 0.068(1) | 0.436(1) | 1 | 3.53(9) |
| Si | 4$a$ | 0.133(1) | 0.750(3) | 0.241(2) | 1 | 0.07(1) |
| S1 | 4$a$ | -0.019(1) | 0.281(3) | 0.644(2) | 1 | 0.07(1) |
| S2 | 4$a$ | 0.126(1) | 0.286(2) | 0.025(2) | 1 | 0.07(1) |
| S3 | 4$a$ | 0.126(1) | 0.492(3) | 0.380(2) | 1 | 0.07(1) |
| S4 | 4$a$ | 0.266(1) | 0.229(2) | 0.644(2) | 1 | 0.07(1) |
| S5 | 4$a$ | 0.378(1) | 0.325(2) | 0.291(3) | 1 | 0.07(1) |
| S6 | 4$a$ | 0.625(1) | 0.551(2) | 0.390(2) | 1 | 0.07(1) |

*Table S2. Structural parameters of Ag$_8$GeS$_6$ at 298 K as obtained by Rietveld refinements and utilizing laboratory X-ray diffraction (Mo Kα radiation). Results for this structure have been previously reported by Ghata et al.* [3]

| Structural information of Ag$_8$GeS$_6$ from X-ray diffraction data at 298 K |||||||
|---|---|---|---|---|---|---|
| Space group: $Pna2_1$; λ (Mo Kα) = 0.7093Å; |||||||
| Lattice parameters: $a$ = 15.147(1) Å, $b$ = 7.4695(5) Å, $c$ = 10.5852(7) Å, |||||||
| $R_{wp}$ = 4.78%; GoF = 1.49 |||||||
| **Atom** | **Wyckoff site** | $x$ | $y$ | $z$ | Occ. | $B_{eq}$ / Å$^2$ |
| Ag1 | 4$a$ | 0.0170(3) | 0.0082(9) | 0.0172(6) | 1 | 3.39(6) |
| Ag2 | 4$a$ | 0.0625(3) | 0.2250(6) | 0.2556(5) | 1 | 3.39(6) |
| Ag3 | 4$a$ | 0.1247(3) | 0.2239(6) | 0.7940(6) | 1 | 3.39(6) |
| Ag4 | 4$a$ | 0.2234(3) | -0.0007(7) | -0.0019(5) | 1 | 3.39(6) |
| Ag5 | 4$a$ | 0.2605(3) | 0.1251(7) | 0.3209(5) | 1 | 3.39(6) |
| Ag6 | 4$a$ | 0.2731(3) | 0.3800(6) | 0.0990(4) | 1 | 3.39(6) |
| Ag7 | 4$a$ | 0.4154(5) | 0.0997(9) | 0.1194(8) | 1 | 3.39(6) |
| Ag8 | 4$a$ | 0.4351(4) | 0.0649(7) | 0.4350(4) | 1 | 3.39(6) |



| | | | | | | |
|---|---|---|---|---|---|---|
| Ge | 4a | 0.1247(3) | 0.7301(7) | 0.2664(7) | 1 | 0.48(7) |
| S1 | 4a | -0.007(1) | 0.271(2) | 0.643(1) | 1 | 0.48(7) |
| S2 | 4a | 0.1218(9) | 0.268(1) | 0.027(1) | 1 | 0.48(7) |
| S3 | 4a | 0.128(1) | 0.471(1) | 0.392(1) | 1 | 0.48(7) |
| S4 | 4a | 0.257(1) | 0.231(1) | 0.638(1) | 1 | 0.48(7) |
| S5 | 4a | 0.3870(9) | 0.323(1) | 0.299(1) | 1 | 0.48(7) |
| S6 | 4a | 0.631(1) | 0.519(1) | 0.400(1) | 1 | 0.48(7) |

*Table S3.* *Structural parameters of Ag$_8$SnS$_6$ at 298 K as obtained by Rietveld refinements and utilizing laboratory X-ray diffraction (Mo Kα radiation).*

| Structural information of Ag$_8$SnS$_6$ from X-ray diffraction data at 298 K | | | | | | |
|---|---|---|---|---|---|---|
| Space group: $Pna2_1$; λ (Mo Kα) = 0.7093Å; | | | | | | |
| Lattice parameters: $a$ = 15.3119(7) Å, $b$ = 7.5542(5) Å, $c$ = 10.7071(5) Å, | | | | | | |
| $R_{wp}$ = 5.36%; GoF = 1.23 | | | | | | |
| **Atom** | **Wyckoff site** | $x$ | $y$ | $z$ | Occ. | $B_{eq}$ / Å$^2$ |
| Ag1 | 4a | 0.0198(3) | 0.1463(9) | 0.0165(6) | 1 | 3.55(6) |
| Ag2 | 4a | 0.0617(3) | 0.2273(6) | 0.2533(8) | 1 | 3.55(6) |
| Ag3 | 4a | 0.1233(3) | 0.2201(7) | 0.8024(9) | 1 | 3.55(6) |
| Ag4 | 4a | 0.2224(3) | 0.0096(7) | 0.0056(9) | 1 | 3.55(6) |
| Ag5 | 4a | 0.2576(4) | 0.1562(7) | 0.3465(8) | 1 | 3.55(6) |
| Ag6 | 4a | 0.2768(3) | 0.3779(6) | 0.1041(8) | 1 | 3.55(6) |
| Ag7 | 4a | 0.4112(3) | 0.0797(7) | 0.1243(8) | 1 | 3.55(6) |
| Ag8 | 4a | 0.4366(4) | 0.0627(7) | 0.4334(8) | 1 | 3.55(6) |
| Sn | 4a | 0.1254(2) | 0.7312(5) | 0.2699(8) | 1 | 0.63(5) |
| S1 | 4a | 0.002(1) | 0.266(2) | 0.639(1) | 1 | 0.63(5) |
| S2 | 4a | 0.1257(9) | 0.278(1) | 0.033(1) | 1 | 0.63(5) |
| S3 | 4a | 0.127(1) | 0.476(1) | 0.397(1) | 1 | 0.63(5) |
| S4 | 4a | 0.247(1) | 0.229(1) | 0.629(1) | 1 | 0.63(5) |
| S5 | 4a | 0.3897(9) | 0.323(1) | 0.296(1) | 1 | 0.63(5) |
| S6 | 4a | 0.631(1) | 0.516(1) | 0.408(1) | 1 | 0.63(5) |

*Table S4.* *Temperature-dependent variations in lattice parameters and unit-cell volumes of Ag$_8$SiS$_6$ from 100 K to 400 K based on Rietveld refinements.*

| Temperature / K | Space group | Lattice parameters / Å | Unit cell volume / Å$^3$ |
|---|---|---|---|
| 103 | $Pna2_1$ | a = 15.009(1) <br> b = 7.4122(7) <br> c = 10.495(1) | 1167.7(2) |



| | | a = 15.016(1) | |
| --- | --- | --- | --- |
| 123 | $Pna2_1$ | b = 7.4155(6) | 1169.4(1) |
| | | c = 10.5019(9) | |
| | | a = 15.019(1) | |
| 133 | $Pna2_1$ | b = 7.4173(7) | 1170.2(1) |
| | | c = 10.5038(8) | |
| | | a = 15.022(1) | |
| 153 | $Pna2_1$ | b = 7.4184(7) | 1171.0(2) |
| | | c = 10.508(1) | |
| | | a = 15.029(1) | |
| 173 | $Pna2_1$ | b = 7.4218(6) | 1172.9(1) |
| | | c = 10.5153(8) | |
| | | a = 15.031(1) | |
| 193 | $Pna2_1$ | b = 7.4240(9) | 1173.6(2) |
| | | c = 10.517(1) | |
| | | a = 15.034(1) | |
| 213 | $Pna2_1$ | b = 7.4252(7) | 1174.6(2) |
| | | c = 10.521(1) | |
| | | a = 15.041(1) | |
| 233 | $Pna2_1$ | b = 7.4279(7) | 1176.1(2) |
| | | c = 10.526(1) | |
| | | a = 15.050(1) | |
| 253 | $Pna2_1$ | b = 7.4320(7) | 1178.2(1) |
| | | c = 10.533(1) | |
| | | a = 15.053(1) | |
| 273 | $Pna2_1$ | b = 7.4346(7) | 1179.2(2) |
| | | c = 10.536(1) | |
| | | a = 15.058(1) | |
| 293 | $Pna2_1$ | b = 7.4355(9) | 1180.1(2) |
| | | c = 10.539(1) | |
| | | a = 15.062(1) | |
| 313 | $Pna2_1$ | b = 7.4376(8) | 1181.2(2) |
| | | c = 10.544(1) | |
| | | a = 15.072(1) | |
| 333 | $Pna2_1$ | b = 7.4422(8) | 1183.4(2) |
| | | c = 10.550(1) | |



| Temperature / K | Space group | Lattice parameters / Å | Unit cell volume / Å³ |
|---|---|---|---|
| 353 | $Pna2_1$ | a = 15.078(1)<br>b = 7.4448(9)<br>c = 10.555(1) | 1184.9(2) |
| 373 | $Pna2_1$ | a = 15.084(1)<br>b = 7.4493(8)<br>c = 10.561(1) | 1186.7(2) |
| 400 | $Pna2_1$ | a = 15.092(1)<br>b = 7.4524(8)<br>c = 10.566(1) | 1188.5(2) |

*Table S5.* *Temperature-dependent variations in lattice parameters and unit-cell volumes of Ag₈GeS₆ from 100 K to 400 K based on Rietveld refinements.*

| Temperature / K | Space group | Lattice parameters / Å | Unit cell volume / Å³ |
|---|---|---|---|
| 103 | $Pna2_1$ | a = 15.110(1)<br>b = 7.4477(5)<br>c = 10.5483(8) | 1187.1(1) |
| 123 | $Pna2_1$ | a = 15.111(1)<br>b = 7.4478(6)<br>c = 10.5494(8) | 1187.2(1) |
| 133 | $Pna2_1$ | a = 15.110(1)<br>b = 7.4480(5)<br>c = 10.5493(7) | 1187.3(1) |
| 153 | $Pna2_1$ | a = 15.115(1)<br>b = 7.4502(6)<br>c = 10.5544(9) | 1188.6(1) |
| 173 | $Pna2_1$ | a = 15.119(1)<br>b = 7.4536(6)<br>c = 10.5582(9) | 1189.8(1) |
| 193 | $Pna2_1$ | a = 15.122(1)<br>b = 7.4547(6)<br>c = 10.5628(9) | 1190.7(1) |
| 213 | $Pna2_1$ | a = 15.129(1)<br>b = 7.4587(7)<br>c = 10.567(1) | 1192.4(1) |



| Temperature / K | Space group | Lattice parameters / Å | Unit cell volume / Å³ |
|---|---|---|---|
| 233 | $Pna2_1$ | a = 15.133(1)<br>b = 7.4607(7)<br>c = 10.571(1) | 1193.6(1) |
| 253 | $Pna2_1$ | a = 15.136(1)<br>b = 7.4656(7)<br>c = 10.5797(9) | 1194.5(1) |
| 273 | $Pna2_1$ | a = 15.141(1)<br>b = 7.4376(8)<br>c = 10.544(1) | 1195.9(1) |
| 298 | $Pna2_1$ | a = 15.147(1)<br>b = 7.4695(5)<br>c = 10.5852(7) | 1197.7(1) |
| 313 | $Pna2_1$ | a = 15.152(1)<br>b = 7.4718(8)<br>c = 10.588(1) | 1198.8(2) |
| 333 | $Pna2_1$ | a = 15.157(2)<br>b = 7.474(1)<br>c = 10.593(1) | 1200.0(3) |
| 353 | $Pna2_1$ | a = 15.162(1)<br>b = 7.4783(8)<br>c = 10.598(1) | 1201.8(2) |
| 373 | $Pna2_1$ | a = 15.169(1)<br>b = 7.4815(8)<br>c = 10.603(1) | 1203.3(2) |
| 400 | $Pna2_1$ | a = 15.169(2)<br>b = 7.483(1)<br>c = 10.606(1) | 1204.1(9) |

*Table S6. Temperature-dependent variations in lattice parameters and unit-cell volumes of $Ag_8SnS_6$ from 100 K to 400 K based on Rietveld refinements.*

| Temperature / K | Space group | Lattice parameters / Å | Unit cell volume / Å³ |
|---|---|---|---|
| 103 | $Pna2_1$ | a = 15.2835 (6)<br>b = 7.5221(3)<br>c = 10.6698(4) | 1226.66(9) |
| | | a = 15.2855(6) | |



| | | | |
|---|---|---|---|
| 123 | $Pna2_1$ | b = 7.5248(3)<br>c = 10.6727(4) | 1227.59(9) |
| 133 | $Pna2_1$ | a = 15.2860(6)<br>b = 7.5253(3)<br>c = 10.6738(4) | 1227.83(9) |
| 153 | $Pna2_1$ | a = 15.2876(9)<br>b = 7.5282(4)<br>c = 10.6765(4) | 1228.7(1) |
| 173 | $Pna2_1$ | a = 15.2889(9)<br>b = 7.5305(4)<br>c = 10.6793(6) | 1229.5(1) |
| 193 | $Pna2_1$ | a = 15.292(9)<br>b = 7.5344(6)<br>c = 10.6838(9) | 1231.0(1) |
| 213 | $Pna2_1$ | a = 15.296(2)<br>b = 7.5389(5)<br>c = 10.6884(7) | 1232.5(1) |
| 233 | $Pna2_1$ | a = 15.299(1)<br>b = 7.5423(5)<br>c = 10.6928(7) | 1233.9(1) |
| 253 | $Pna2_1$ | a = 15.304(1)<br>b = 7.5467(5)<br>c = 10.6986(7) | 1235.6(1) |
| 273 | $Pna2_1$ | a = 15.3103(8)<br>b = 7.5510(4)<br>c = 10.7034(6) | 1237.4(1) |
| 293 | $Pna2_1$ | a = 15.311(1)<br>b = 7.5534(8)<br>c = 10.706(1) | 1238.2(2) |
| 313 | $Pna2_1$ | a = 15.317(1)<br>b = 7.5586(1)<br>c = 10.7134(1) | 1240.4(1) |
| 333 | $Pna2_1$ | a = 15.322(1)<br>b = 7.5625(6)<br>c = 10.7174(9) | 1241.8(1) |
| | | a = 15.325(1) | |



| 353 | $Pna2_1$ | b = 7.5661(7) <br> c = 10.7228(9) | 1243.3(1) |
| 373 | $Pna2_1$ | a = 15.328(1) <br> b = 7.5696(9) <br> c = 10.726(1) | 1244.5(2) |
| 400 | $Pna2_1$ | a = 15.332(2) <br> b = 7.575(1) <br> c = 10.733(1) | 1246.6(3) |

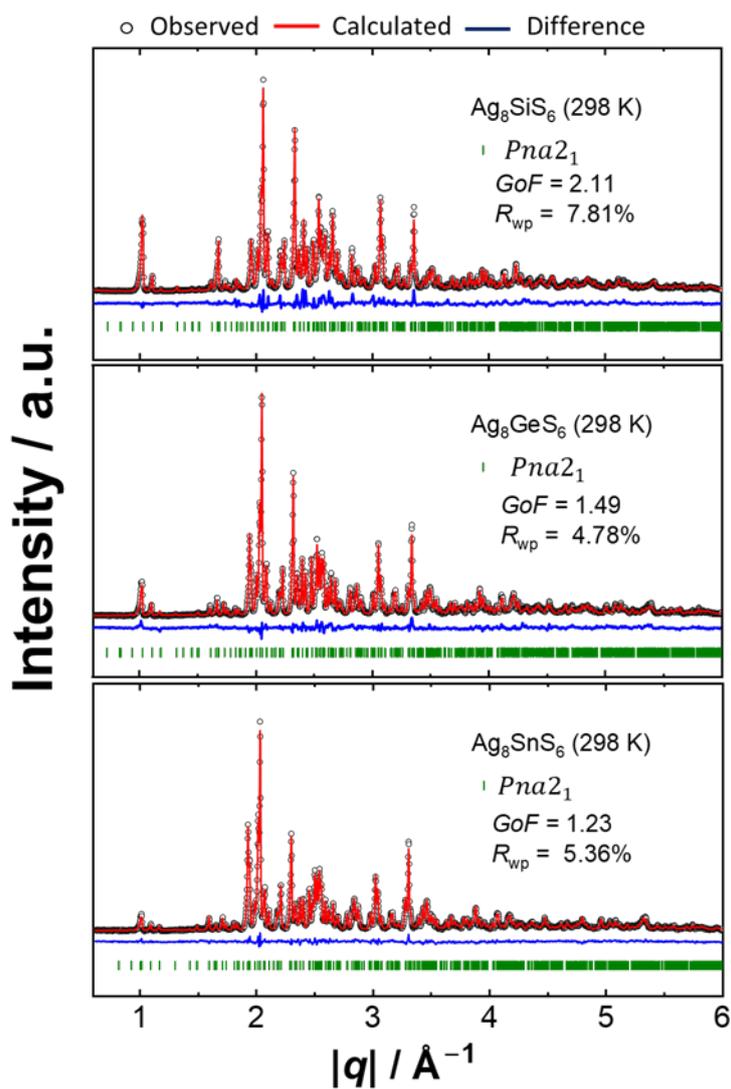

*Figure S2.* X-ray diffraction patterns and corresponding Rietveld refinement results of $Ag_8TS_6$ (T = Si, Ge, Sn) at 298 K.



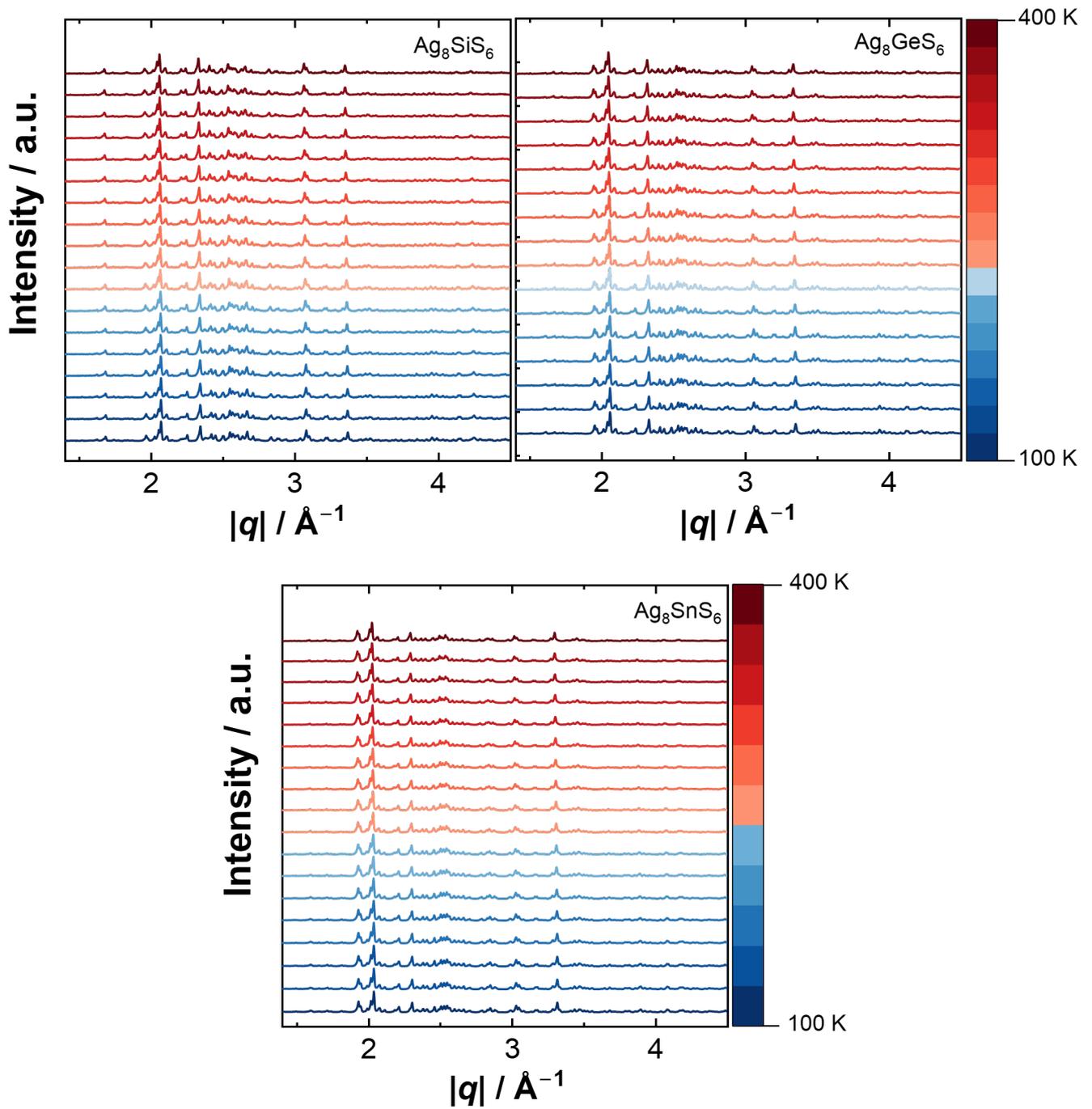

***Figure S3.*** *Temperature-dependent X-ray diffraction patterns of Ag$_8$TS$_6$ (T = Si, Ge, Sn) measured over the temperature range of 100 K to 400 K.*



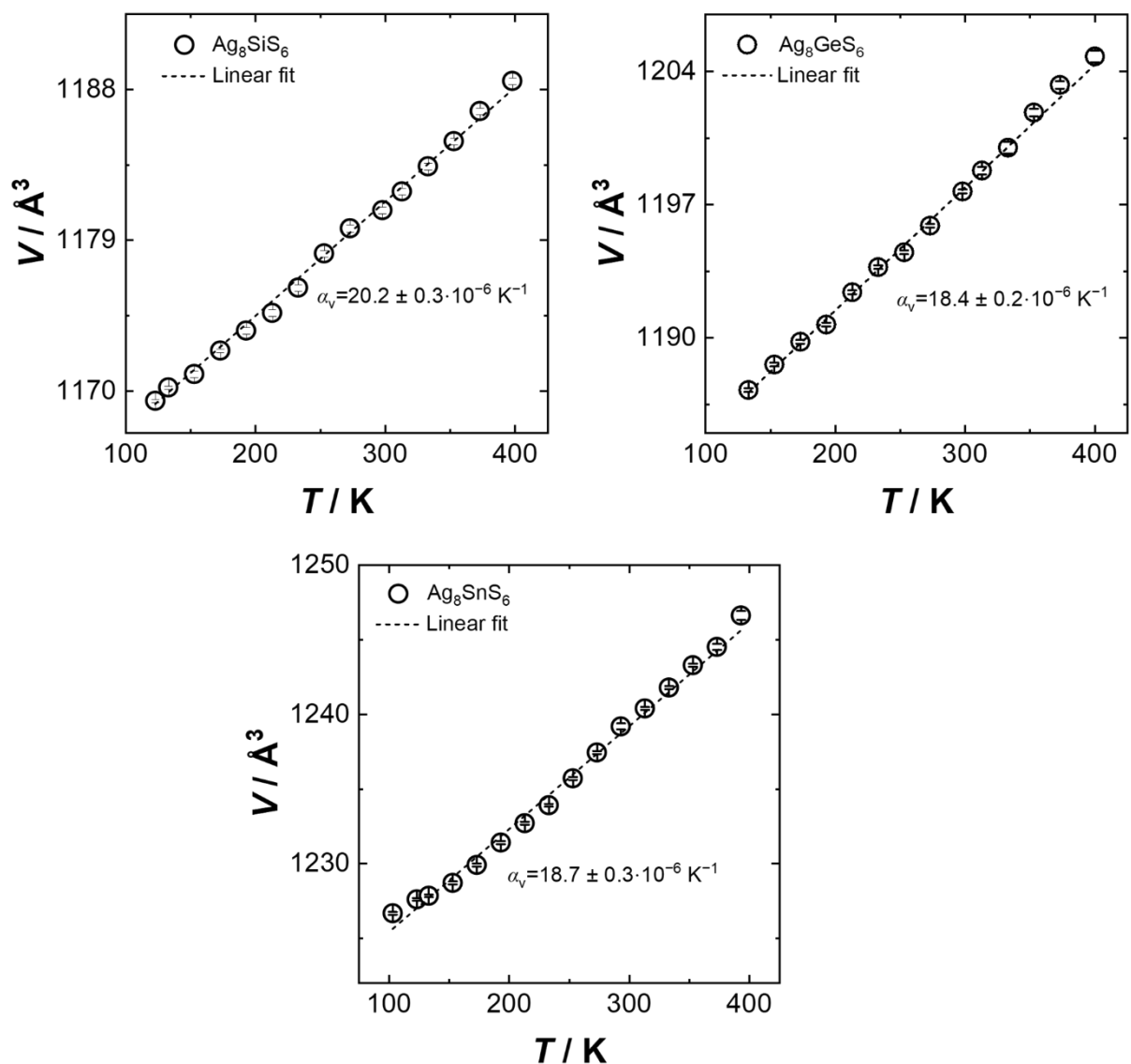

*Figure S4.* Unit cell volume change of $Ag_8TS_6$ (T = Si, Ge, Sn) with temperature; the thermal volume expansion coefficient can be determined from the slopes of the plots.

**Section S2: Computational details – Stability and Bonding analysis**

Computed lattice parameters *a*, *b*, and *c* for all the structures presented in **Figure S1** are reported in **Table S7**. Here, our computed lattice parameters show only ~2% overestimation with respect to experimental values, which lies within the typical error range of DFT calculations and ensures the robustness of our computational approach. In addition, a slight decrease in the volume is observed when we move from Sn to Ge and Si, consistent with the expected trend of decreasing atomic radii from bottom to top within the groups.



*Table S7. Computational lattice parameters for Ag$_8$TS$_6$ (T = Si, Ge, and Sn) compared with experimental measurements obtained by Rietveld refinements at 298 K*

| Unit cell | Ag$_8$SiS$_6$ | | Ag$_8$GeS$_6$ | | Ag$_8$SnS$_6$ (RT) | | Ag$_8$SnS$_6$ (LT) | |
|---|---|---|---|---|---|---|---|---|
| | | Exp* | | Exp* | | Exp* | | Exp** |
| a (Å) | 15.16 | 15.058 | 15.27 | 15.147 | 15.46 | 15.3119 | 7.78 | 7.66 |
| b (Å) | 7.59 | 7.4355 | 7.62 | 7.4695 | 7.70 | 7.5542 | 7.67 | 7.54 |
| c (Å) | 10.69 | 10.5415 | 10.72 | 10.5852 | 10.85 | 10.7071 | 10.86 | 10.63 |
| V (Å$^3$) | 1228.61 | 1180.266 | 1246.84 | 1197.615 | 1291.25 | 1238.481 | 648.10 | 614.15 |
| V/Z (Å$^3$) | 307.15 | 295.0665 | 311.71 | 299.404 | 322.81 | 309.6203 | 324.05 | 307.08 |

*Our Experimental values **Experimental data from Slade's work [4]

Based on the optimised structures, we analysed the material's stability with harmonic phonon calculations. The phonon band structures of all studied argyrodites do not show any imaginary modes, as we show in **Figure 4** and **Figures S5**.

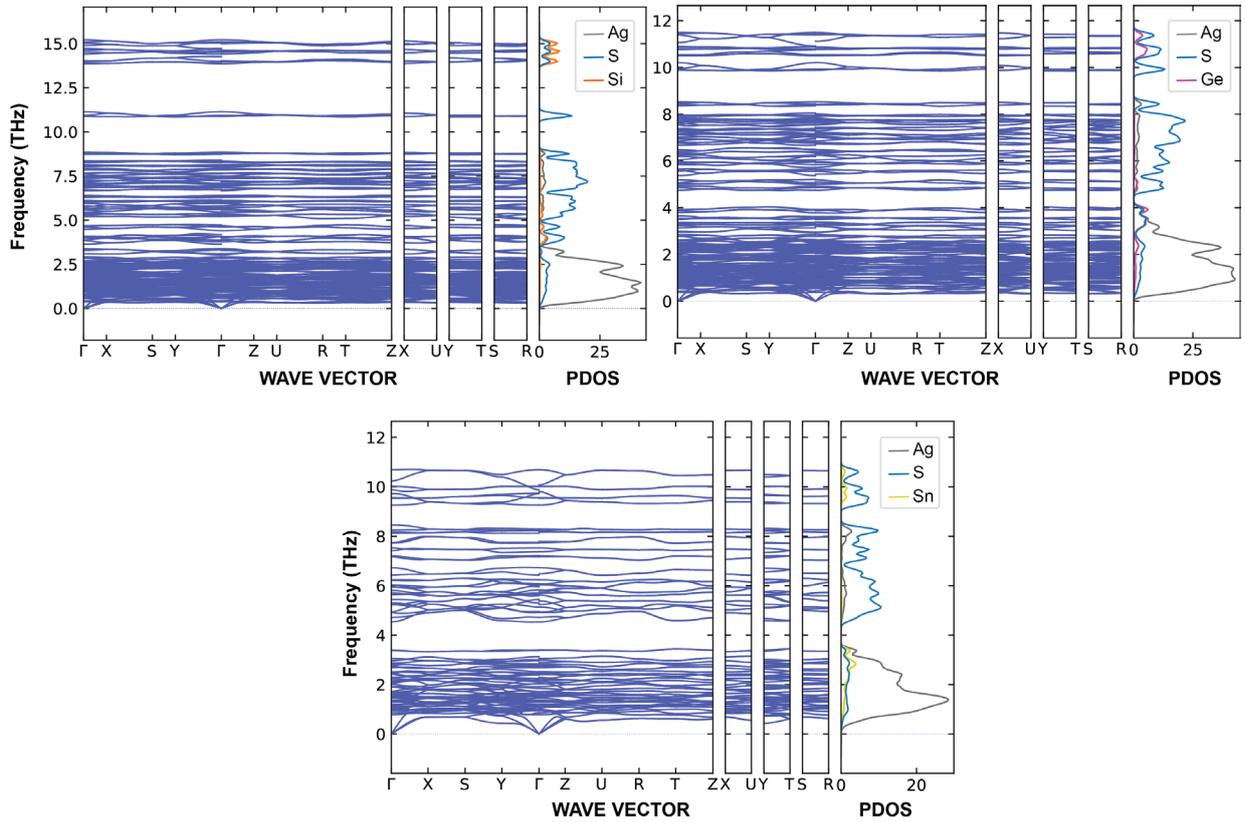

*Figure S5. Phonon dispersion curves for the Ag$_8$SiS$_6$, Ag$_8$GeS$_6$, and low-temperature phase of Ag$_8$SnS$_6$ canfieldite.*



**Bonding analysis**

From our automated bonding analysis, we present coordination environments, Wyckoff positions, ICOHPs, and two-centre ICOBIs per bond for all our argyrodite compounds $Ag_8TS_6$ ($T$= Si, Ge, and Sn at room and low temperature) in **Figures S6-S9**. We considered ICOHPs and ICOBIs for Ag–Ag bonds less than 3Å.



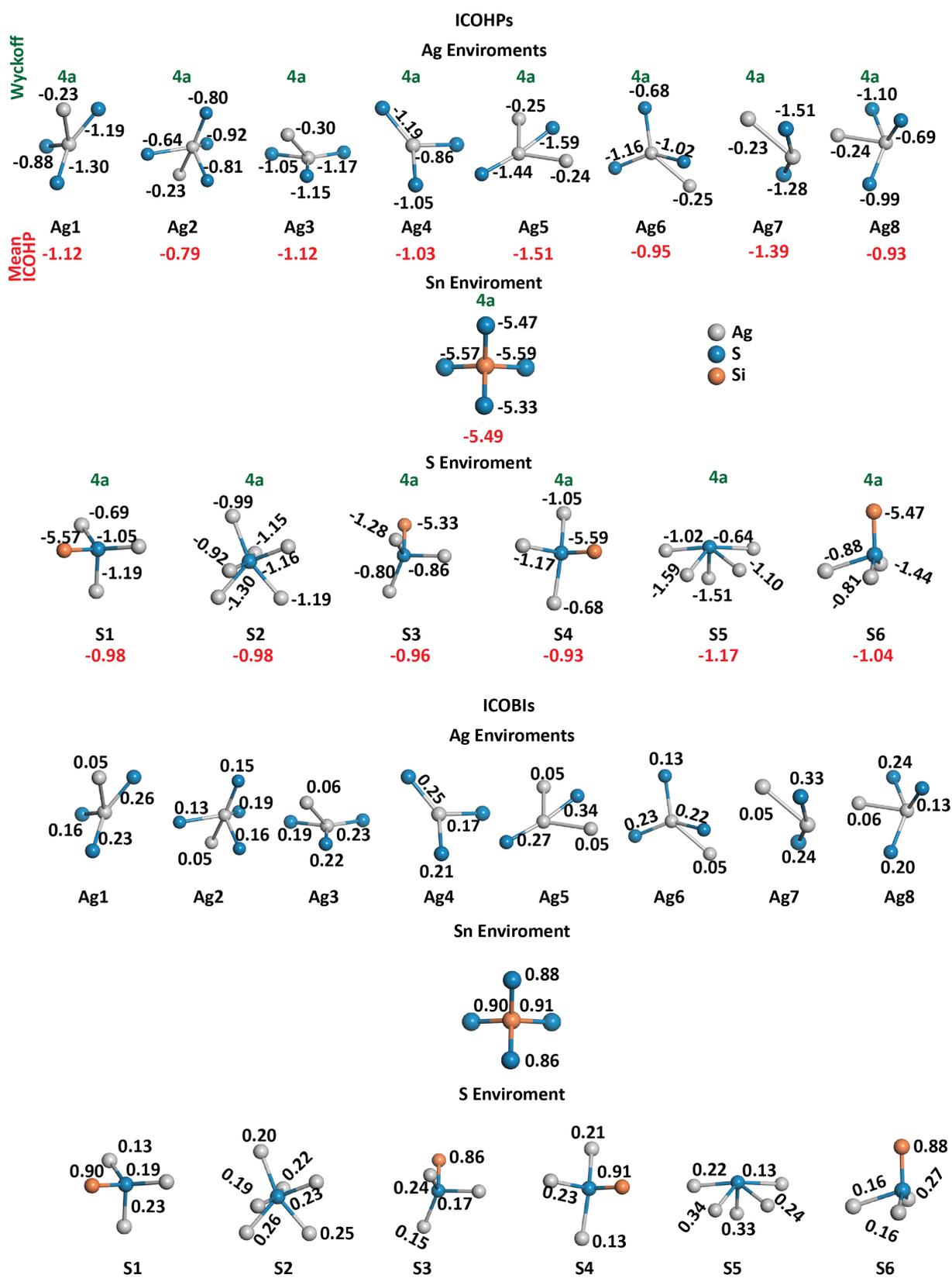

*Figure S6.* ICOHPs and two-center ICOBIs for $Ag_8SiS_6$ structure based on the different bonds and coordination environments.



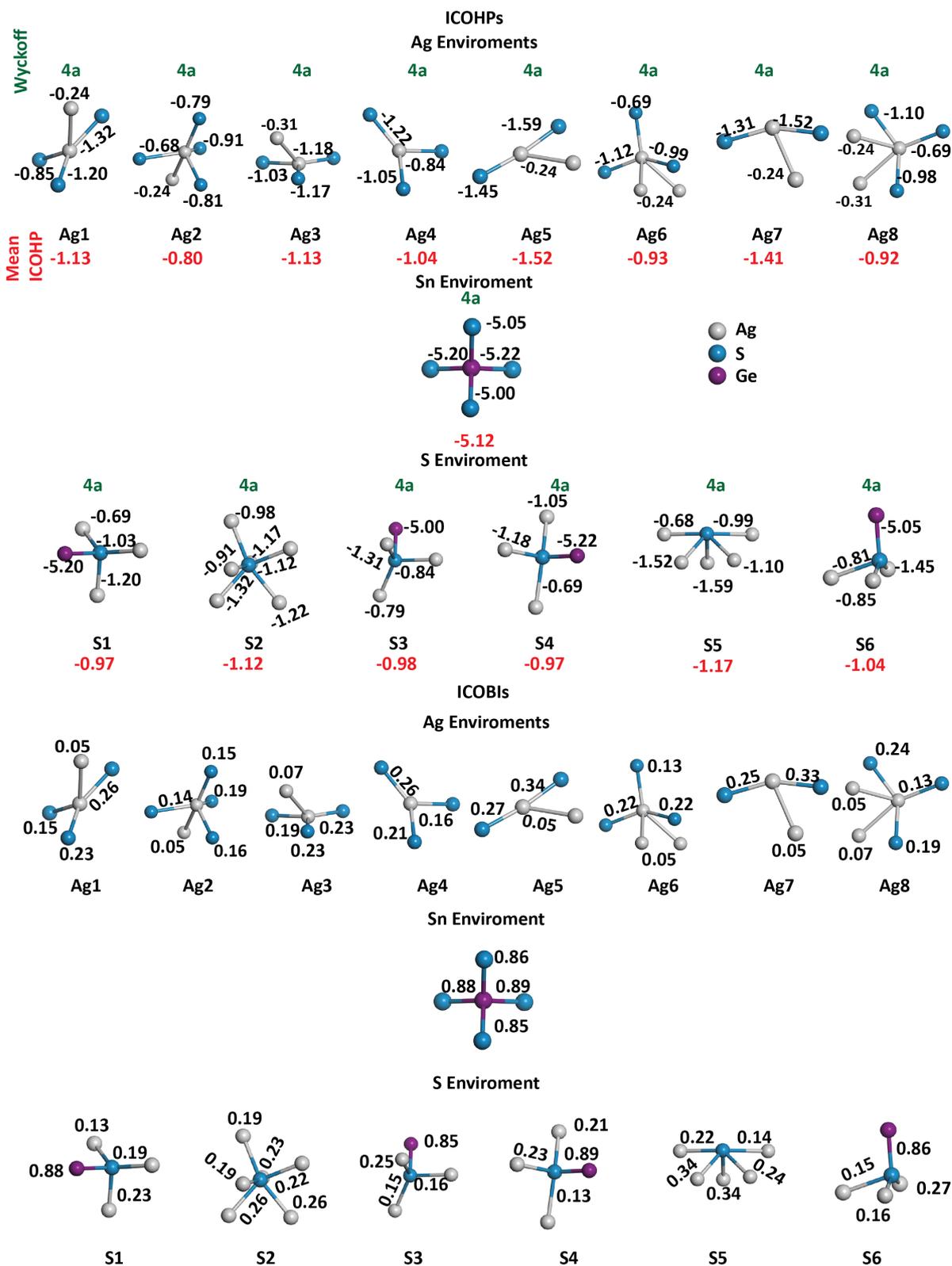

***Figure S7***. *ICOHPs and two-centre ICOBIs for Ag$_8$GeS$_6$ structure based on the different bonds and coordination environments.*



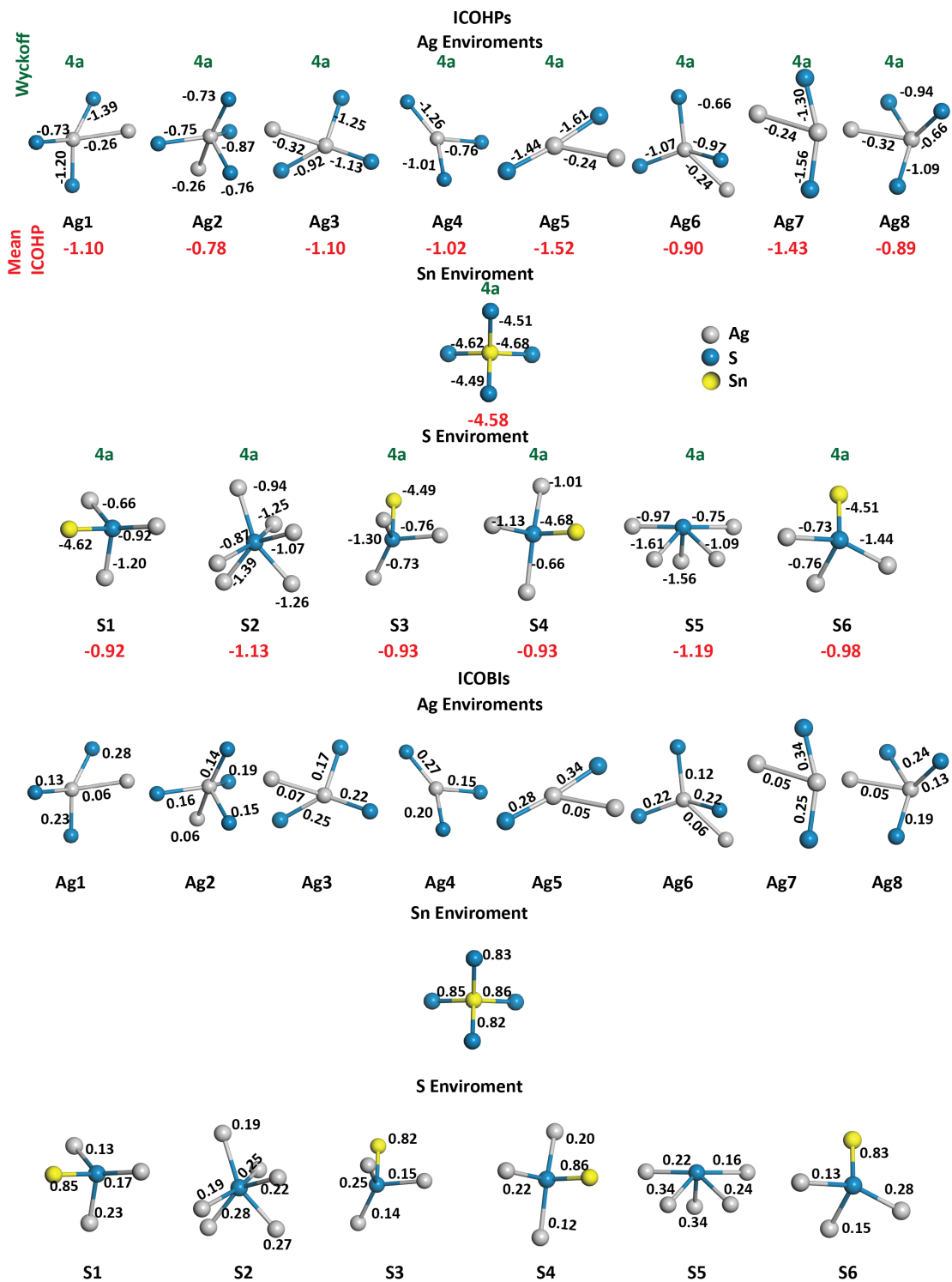

*Figure S8*. Local coordination environments, including ICOHPs and ICOBIs per bond for the canfieldite $Ag_8SnS_6$ at room temperature.



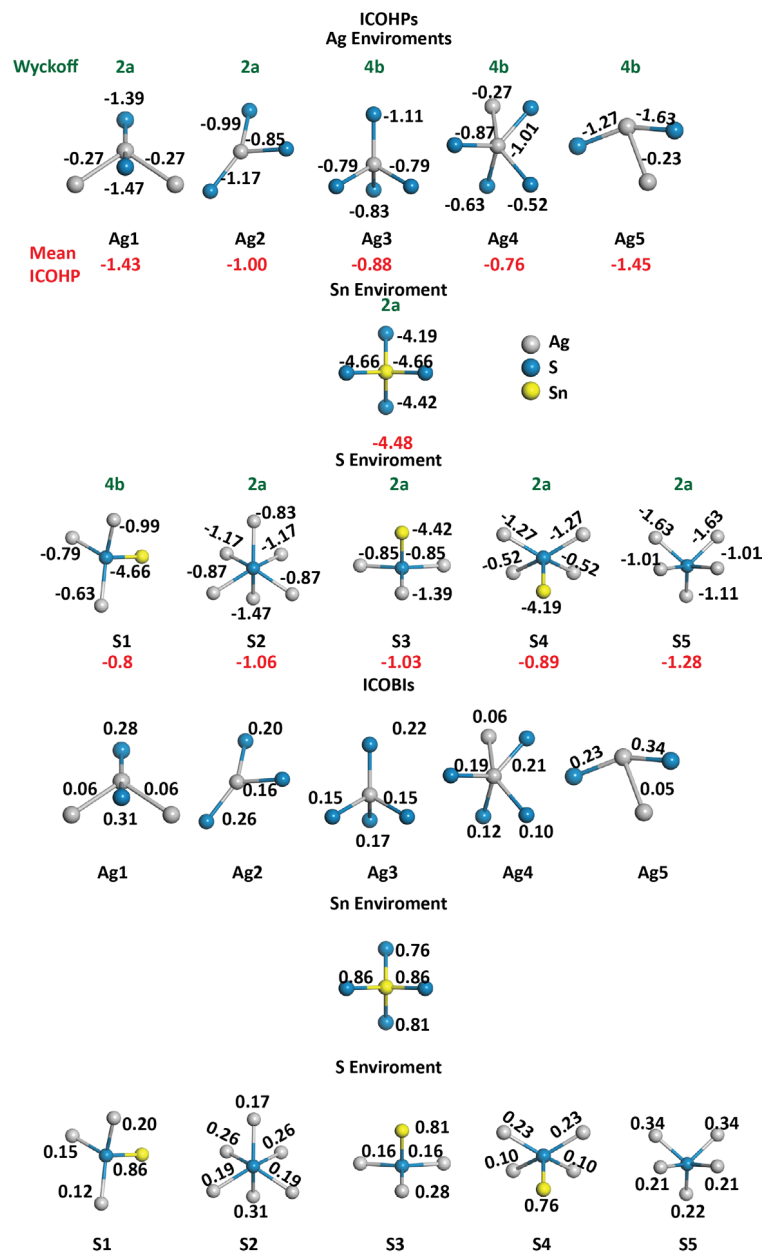

***Figure S9.*** *ICOHPs and two-centre ICOBIs for low-temperature Ag$_8$SnS$_6$ based on the different bonds and coordination environments.*

***Table S8.*** *Atomic positions and inequivalent site fractional coordinates for (Pna2$_1$) Ag$_8$SiS$_6$.*

| Atom | POSCAR Position | Wyckoff Positions | x | y | z | Coordination Environment |
|---|---|---|---|---|---|---|
| Ag1 | Ag5 | 4a | 0.014 | 0.021 | 0.032 | Trigonal Planar |
| Ag2 | Ag9 | 4a | 0.066 | 0.242 | 0.252 | Tetrahedral |
| Ag3 | Ag13 | 4a | 0.129 | 0.284 | 0.783 | Trigonal Planar |
| Ag4 | Ag17 | 4a | 0.228 | 0.007 | 0.009 | Trigonal Planar |
| Ag5 | Ag21 | 4a | 0.261 | 0.125 | 0.329 | Linear |
| Ag6 | Ag25 | 4a | 0.262 | 0.383 | 0.122 | Trigonal Planar |
| Ag7 | Ag29 | 4a | 0.410 | 0.130 | 0.102 | Linear |



| Atom | POSCAR Position | Wyckoff Positions | x | y | z | Coordination Environment |
|---|---|---|---|---|---|---|
| Ag8 | Ag33 | 4a | 0.436 | 0.071 | 0.440 | Triangular Non-coplanar |
| **Si** | Si1 | 4a | 0.123 | 0.742 | 0.267 | Tetrahedral |
| S1 | S37 | 4a | 0.991 | 0.266 | 0.648 | Tetrahedral |
| S2 | S41 | 4a | 0.123 | 0.268 | 0.019 | Trigonal Prismatic |
| S2 | S45 | 4a | 0.120 | 0.513 | 0.388 | Tetrahedral |
| S4 | S49 | 4a | 0.263 | 0.232 | 0.649 | Tetrahedral |
| S5 | S53 | 4a | 0.387 | 0.316 | 0.287 | Square Pyramidal |
| S6 | S57 | 4a | 0.625 | 0.527 | 0.385 | See-saw like |

*Table S9. Atomic positions and inequivalent site fractional coordinates for (Pna2$_1$) Ag$_8$GeS$_6$*

| Atom | POSCAR Position | Wyckoff Positions | x | y | z | Coordination Environment |
|---|---|---|---|---|---|---|
| Ag1 | Ag1 | 4a | 0.015 | 0.022 | 0.033 | Trigonal Planar |
| Ag2 | Ag5 | 4a | 0.064 | 0.24 | 0.252 | Tetrahedral |
| Ag3 | Ag9 | 4a | 0.129 | 0.282 | 0.785 | Trigonal Planar |
| Ag4 | Ag13 | 4a | 0.228 | 0.009 | 0.011 | Trigonal Planar |
| Ag5 | Ag17 | 4a | 0.261 | 0.130 | 0.329 | Linear |
| Ag6 | Ag21 | 4a | 0.266 | 0.382 | 0.117 | Trigonal Planar |
| Ag7 | Ag25 | 4a | 0.409 | 0.124 | 0.104 | Linear |
| Ag8 | Ag29 | 4a | 0.436 | 0.07 | 0.439 | Triangular Non-coplanar |
| **Ge** | Ge33 | 4a | 0.124 | 0.740 | 0.267 | Tetrahedral |
| S1 | S37 | 4a | 0.995 | 0.266 | 0.643 | Tetrahedral |
| S2 | S41 | 4a | 0.123 | 0.269 | 0.021 | Trigonal Prismatic |
| S3 | S45 | 4a | 0.120 | 0.503 | 0.393 | Tetrahedral |
| S4 | S49 | 4a | 0.258 | 0.23 | 0.644 | Tetrahedral |
| S5 | S53 | 4a | 0.387 | 0.315 | 0.286 | Square Pyramidal |
| S6 | S57 | 4a | 0.626 | 0.52 | 0.39 | See-saw like |

*Table S10. Atomic positions and inequivalent site fractional coordinates for (Pna2$_1$) Ag$_8$SnS$_6$*

| Atom | POSCAR Position | Wyckoff Positions | x | y | z | Coordination Environment |
|---|---|---|---|---|---|---|
| Ag1 | Ag1 | 4a | 0.021 | 0.022 | 0.033 | Trigonal Planar |
| Ag2 | Ag5 | 4a | 0.060 | 0.239 | 0.251 | Tetrahedral |
| Ag3 | Ag9 | 4a | 0.129 | 0.279 | 0.789 | Trigonal Planar |
| Ag4 | Ag13 | 4a | 0.227 | 0.013 | 0.016 | Trigonal Planar |
| Ag5 | Ag17 | 4a | 0.261 | 0.138 | 0.332 | Linear |
| Ag6 | Ag21 | 4a | 0.270 | 0.380 | 0.111 | Triangular Non-coplanar |
| Ag7 | Ag25 | 4a | 0.408 | 0.114 | 0.108 | Linear |
| Ag8 | Ag29 | 4a | 0.436 | 0.069 | 0.436 | Triangular Non-coplanar |
| **Sn** | Sn33 | 4a | 0.125 | 0.739 | 0.266 | Tetrahedral |
| S1 | S37 | 4a | 0.001 | 0.267 | 0.634 | Tetrahedral |
| S2 | S41 | 4a | 0.124 | 0.27 | 0.022 | Trigonal Prismatic |
| S3 | S45 | 4a | 0.121 | 0.488 | 0.4 | Tetrahedral |



| | | | | | | |
|---|---|---|---|---|---|---|
| S4 | S49 | 4a | 0.249 | 0.228 | 0.636 | Tetrahedral |
| S5 | S53 | 4a | 0.387 | 0.310 | 0.283 | Square Pyramidal |
| S6 | S57 | 4a | 0.626 | 0.506 | 0.396 | See-saw like |

*Table S11.* *Atomic positions and inequivalent site fractional coordinates for LT (Pmn21)* $Ag_8SnS_6$

| Atom | POSCAR Position | Wyckoff Positions | x | y | z | Coordination Environment |
|---|---|---|---|---|---|---|
| Ag1 | Ag1 | 2a | 0 | 0.313 | 0.186 | Linear |
| Ag2 | Ag3 | 2a | 0.202 | 0.485 | 0.389 | Trigonal Planar |
| Ag3 | Ag7 | 4b | 0 | 0.385 | 0.620 | Tetrahedral |
| Ag4 | Ag9 | 4b | 0.292 | 0.113 | 0.287 | Tetrahedral |
| Ag5 | Ag13 | 4b | 0.273 | 0.153 | 0.014 | Linear |
| **Sn** | Sn17 | 2a | 0 | 0.754 | 0.133 | Tetrahedral |
| S1 | S19 | 4b | 0.242 | 0.24 | 0.757 | Tetrahedral |
| S2 | S23 | 2a | 0 | 0.217 | 0.401 | Octahedral |
| S3 | S25 | 2a | 0 | 0.497 | 0.002 | Tetrahedral |
| S4 | S27 | 2a | 0 | 0.989 | 0.982 | Square Pyramidal |
| S5 | S29 | 2a | 0 | 0.720 | 0.640 | Square Pyramidal |

**Multi-center bonding analysis**

The (Integrated) Crystal Orbital Bond Index (ICOBI) in LOBSTER[5,6] can be a valuable tool to analyze unusual bonding phenomena, as the two-center ICOBI corresponds to the bond order (BO)[6]. Plotting the two-center ICOBI against the bond length leads to the following outcome in **Figure S10**.



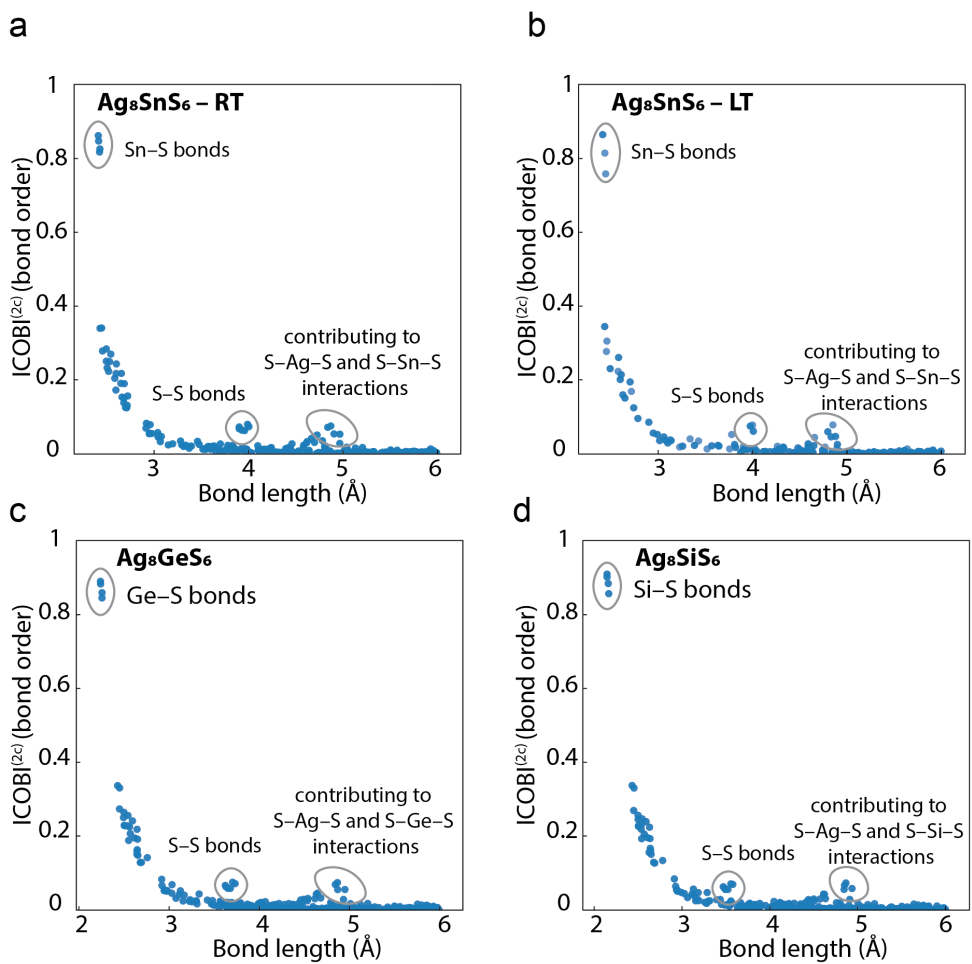

*Figure S10. Two-center ICOBI vs. bond length plot of $Ag_8TS_6$ (T = Sn, Ge, Si).*

All phases of $Ag_8SnS_6$, $Ag_8GeS_6$, and $Ag_8SiS_6$ exhibit $TS_4$ tetrahedra with an ICOBI (or BO) of almost one for each TS bond. These tetrahedra also show stronger S—S bonds. They form the covalent backbone of the structures. Then again, all four compounds display a bunch of unusually strong bonds in the range of 4.5 to 5.5 Å, which are suspected of contributing to multi-center interactions.

Plotting the ICOBI$^{(3c)}$ against the bond angle (**Figure 3** in the main text) and the distance of the terminating atoms (**Figure S11**) reveals almost the same correlation.



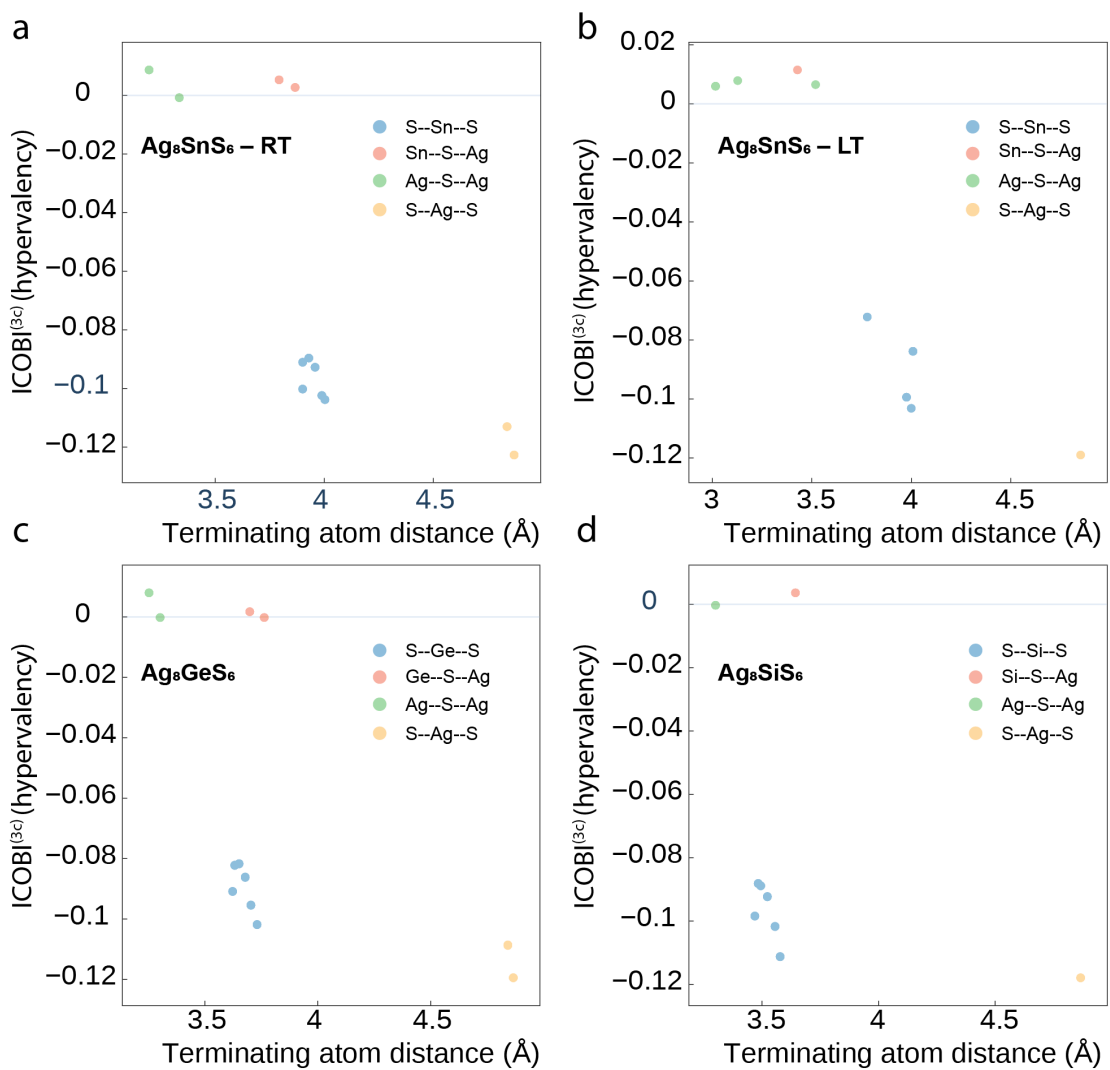

***Figure S11.*** *Three-center ICOBI vs. bond angle plot of Ag$_8$TS$_6$ (T = Sn, Ge, Si).*

The bonding situation involving Ag and S resembles the one in the phase-change material [NaCl] GeTe.[7] In GeTe, it was found that the Te—Ge—Te bonds show an ICOBI$^{(3c)}$ value of around –0.1, while the Ge—Te—Ge ICOBI$^{(3c)}$ is exactly zero. The peculiar bonding situation in GeTe can be explained by constructively interfering orbital contributions for Te—Ge—Te and destructively interfering orbital contributions for Ge—Te—Ge (cf. Figure. 3 in [7]). A similar situation is found for the argyrodite compounds.

For example, in Ag$_8$SnS$_6$ (RT), the S—Ag—S bonds have dominating orbital contributions of S(3p$_y$)—Ag(5s)—S(3p$_x$) (−0.02026), S(3p$_x$)—Ag(5s)—S(3p$_y$) (−0.01288), S(3p$_x$)—Ag(5s)—S(3s) (−0.01150), and S(3p$_x$)—Ag(5s)—S(3p$_x$) (−0.02862), while the Ag—S—Ag



bond's leading orbital contributions Ag(5s)—S(3p$_y$)—Ag(5s) (+0.02194) and Ag(5s)—S(3p$_x$)—Ag(5s) (−0.01536) are almost canceling each other out.

In Ag$_8$SnS$_6$ (LT), the Ag—S—Ag orbital contributions are all smaller than ±0.01, while the dominating S—Ag—S orbital contributions are S(3p$_z$)—Ag(5s)—S(3s) (−0.01292), S(3p$_z$)—Ag(5s)—S(3p$_y$) (−0.03628) and S(3p$_z$)—Ag(5s)—S(3p$_z$) (−0.03566).

In Ag$_8$GeS$_6$, the S—Ag—S leading orbital contributions are S(3p$_y$)—Ag(5s)—S(3p$_x$) (−0.02136), S(3p$_x$)—Ag(5s)—S(3p$_y$) (−0.01276), S(3p$_x$)—Ag(5s)—S(3s) (−0.01168), and S(3p$_x$)—Ag(5s)—S(3p$_x$) (−0.02612), while the Ag—S—Ag orbital contributions are Ag(5s)—S(3p$_y$)—Ag(5s) (+0.01798) and Ag(5s)—S(3p$_x$)—Ag(5s) (−0.01298).

Finally, in Ag$_8$SiS$_6$, the S—Ag—S leading orbital contributions are S(3p$_y$)—Ag(5s)—S(3p$_x$) (−0.02174), S(3p$_x$)—Ag(5s)—S(3p$_y$) (−0.01346), S(3p$_x$)—Ag(5s)—S(3s) (−0.01142), and S(3p$_x$)—Ag(5s)—S(3p$_x$) (−0.02488), while the Ag—S—Ag orbital contributions are Ag(5s)—S(3p$_y$)—Ag(5s) (+0.01598), Ag(5s)—S(3p$_z$)—Ag(5s) (−0.00786) and Ag(5s)—S(3p$_x$)—Ag(5s) (−0.00840).

Except for the expected difference in the distance of the terminating atoms of the tetrahedral bonds because of the different metal types, there is almost no difference in the three-center bonds of Ag$_8$SnS$_6$, Ag$_8$GeS$_6,$ and Ag$_8$SiS$_6$.

### Section S3: Thermal transport

The total thermal conductivity $\kappa$ is calculated from the measured thermal diffusivity $D$ following the equation $\kappa = D \cdot C_p \cdot \rho$, where $\rho$ represents the geometrical density and $C_p$ is the isobaric heat-capacity. The isobaric heat-capacity was approximated using isochoric heat capacities derived from density-functional theory simulations. The total thermal conductivity comprises contributions from both the lattice thermal conductivity ($\kappa_{lat}$) and the electronic thermal conductivity ($\kappa_e$). The electronic thermal conductivity ($\kappa_e$) can be estimated using the Wiedemann-Franz law, $\kappa_e = L \cdot \sigma \cdot T$, where $L$ is the Lorenz number, $\sigma$ is the electrical conductivity, and $T$ is the temperature.[8] For the Ag$_8$$T$S$_6$ ($T$ = Si, Ge, Sn) argyrodites, the electrical conductivity ($\sigma$) is below the detection limit (minimum measurable value: 0.05 S/cm) of our four-probe measurement setup (SBA 458 instrument), making it unmeasurable. Therefore, the total thermal conductivity ($\kappa$) is assumed to be equivalent to the lattice thermal conductivity ($\kappa_{lat}$) for these systems. Thermal diffusivity ($D$) (**Figure S12**) and total thermal conductivity ($\kappa$) across the three compositions show similar results, with no significant variations observed.

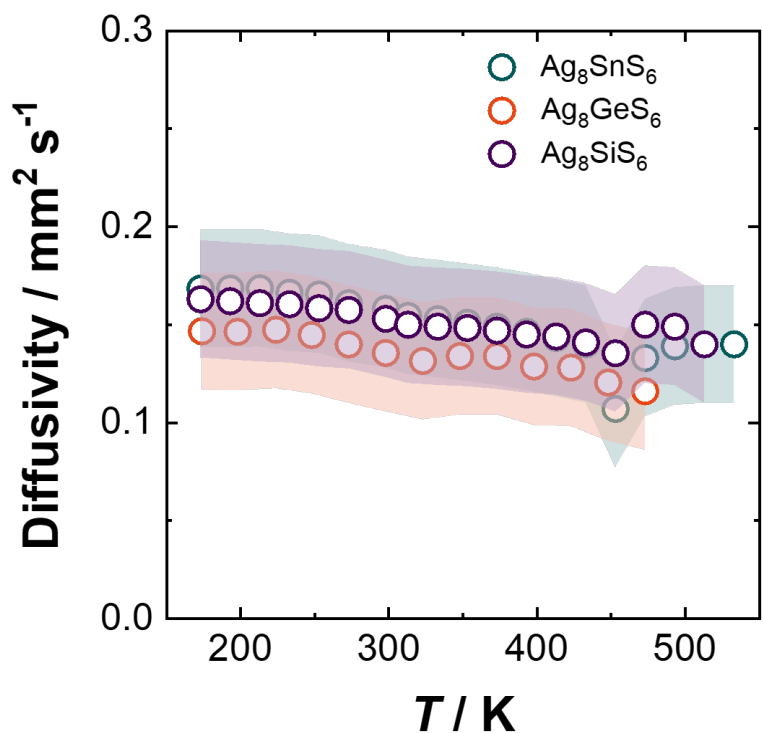

*Figure S12.* Temperature-dependent thermal diffusivity (D) of $Ag_8TS_6$ (T = Si, Ge, Sn).

**Section S4: Synthesis and X-ray diffraction analysis of blocking electrode $RbAg_4I_5$**

$RbAg_4I_5$ has previously been reported as an effective $Ag^+$ ion-conducting and electron-blocking electrode.[9] In this study, it is utilized to prevent electronic interference and enable accurate measurement of ionic conductivity of our Ag-based argyrodites. $RbAg_4I_5$ was synthesized via mechanochemical ball milling process. Stoichiometric amounts of RbI (Thermo Scientific, 99.8%) and AgI (Thermo Scientific, 99 %) were weighed inside an argon-filled glovebox under dark conditions and pre-mixed by hand grinding. The mixture was transferred into 80 mL zirconia ball milling cups, along with 5 mm diameter milling media (10:1 ball to reactant mixture mass ratio) and milled for 72 cycles at 400 rpm (10 minutes of milling followed by 10 minutes of rest per cycle). Upon completing the 72 cycles, the ball-milling cups were opened inside the glovebox, and the samples were taken and hand-ground in an agate mortar.

The X-ray diffraction pattern of powdered $RbAg_4I_5$ was measured and analyzed by Rietveld refinement using the TOPAS-Academic V7 software package[1], confirming a cubic structure with space group ($P4_132$) at 298 K (**Figure S13**). The observed phase fully accounts for the diffraction pattern and the refined lattice parameter, $a$ = 11.2493(3) Å, agrees well with literature values,[10] verifying the successful synthesis and phase purity of the compound.

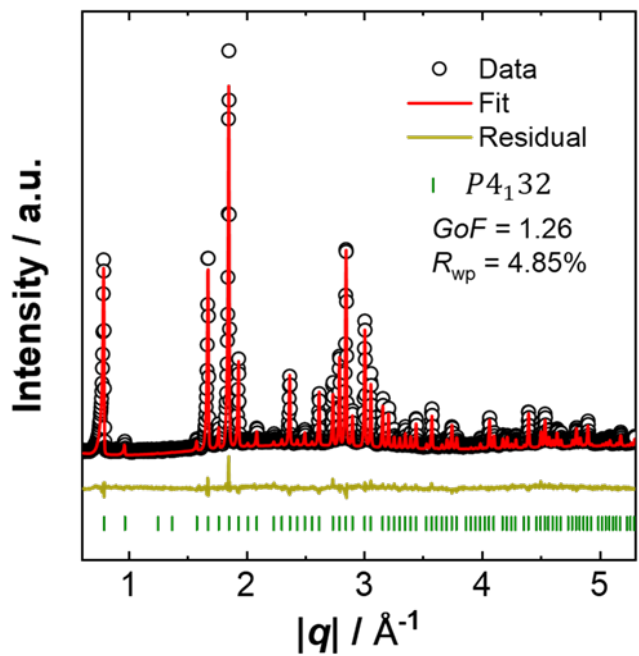

*Figure S13. X-ray diffraction patterns and corresponding Rietveld refinement results of RbAg$_4$I$_5$ at 298 K.*

**Section S5: DC polarization and ion transport**

Ag$_8T$S$_6$ ($T$ = Si, Ge, Sn) exhibit very low electronic conductivity, as confirmed by electronic DC measurements (**Figure S14**). The impedance results of the electrode material RbAg$_4$I$_5$ and all Ag$_8T$S$_6$ ($T$ = Si, Ge, Sn) argyrodites were analyzed using the RelaxIS 3 software package. The Nyquist plot of RbAg$_4$I$_5$ at 233 K is presented in **Figure S15a**. These measurements, which were also employed in our previous work,[3] were reanalyzed here for completeness and comparison with the related compounds Ag$_8$SiS$_6$ and Ag$_8$SnS$_6$. The impedance response of RbAg$_4$I$_5$ is characterized by an ohmic resistance (x-axis intercept), followed by a straight-line indicative of capacitive behaviour. The ionic conductivity of RbAg$_4$I$_5$ was found to be 173 ± 5 mS / cm, with an activation energy of 0.09 ± 0.02 eV, as illustrated by the Arrhenius plot (**Figure S15b**). The impedance responses of Ag$_8T$S$_6$ ($T$ = Si, Ge, Sn) measured with blocking electrodes (thickness $h$ = 0.08 cm) at 233 K are shown in **Figure 8a**. The spectra exhibit a semicircle at high frequencies and a tail at low frequencies. The high-frequency process is modelled using an equivalent circuit consisting of a resistor ($R$) in parallel with a CPE.

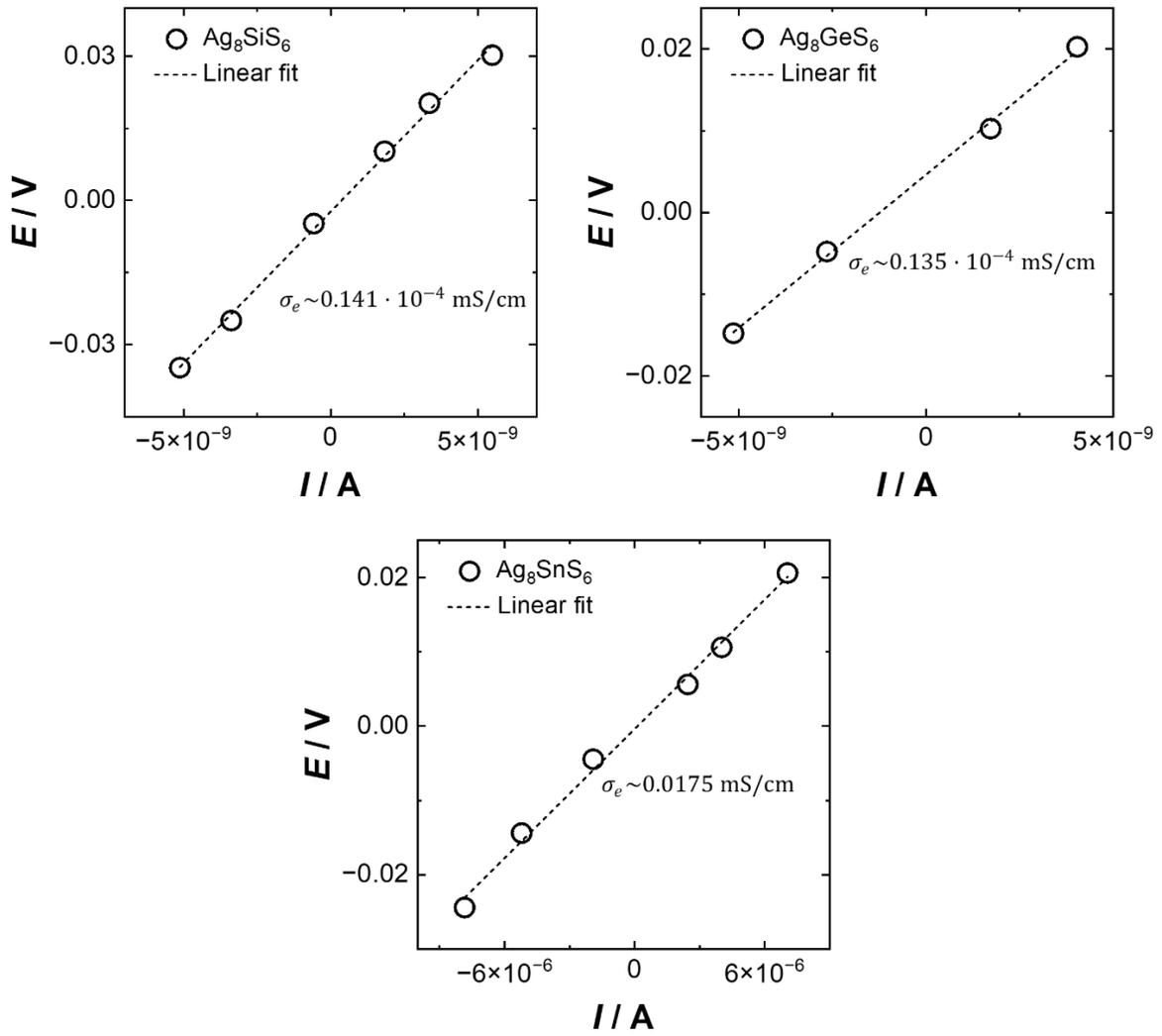

*Figure S14.* The applied voltage versus current response of $Ag_8TS_6$ (T = Si, Ge, Sn), obtained from DC polarization measurements, is used to determine the electronic conductivity ($\sigma_e$) value.

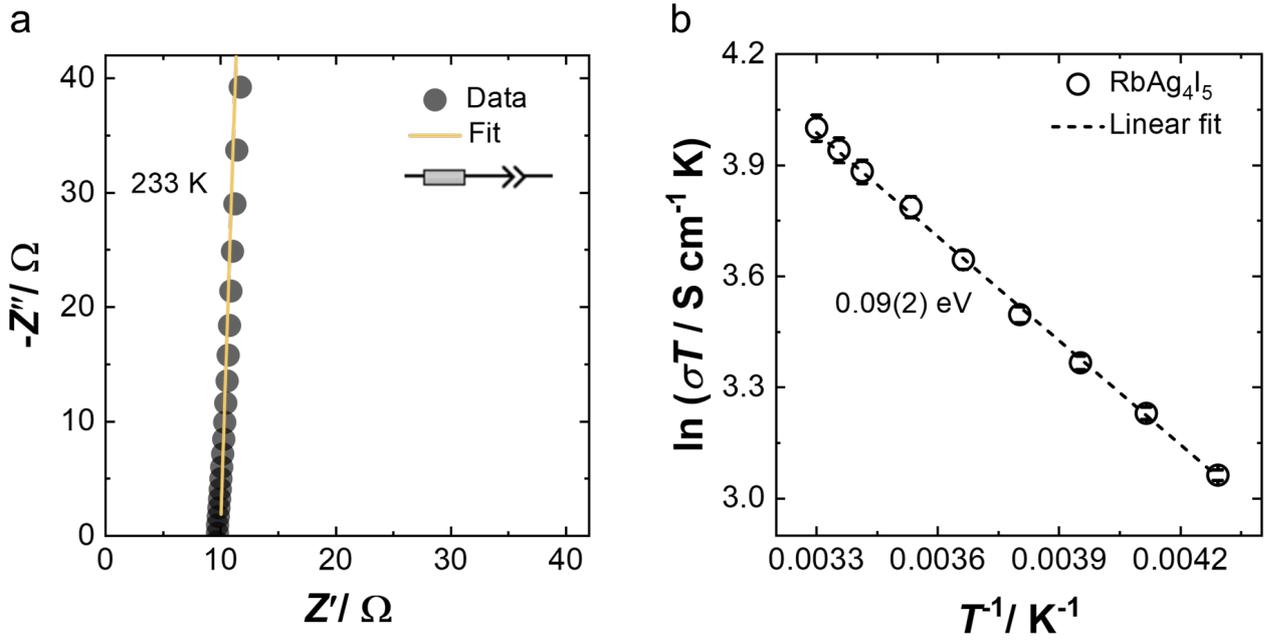

*Figure S15.* a) Nyquist plots of RbAg$_4$I$_5$ at 233 K measured within the frequency range from 5 MHz to 1 Hz. The inset of the plot shows the equivalent circuit used to fit the data. b) Arrhenius plot of measured conductivities of RbAg$_4$I$_5$, indicating an activation energy of 0.09 eV for ion transport.

To interpret the origin of the high-frequency resistance, the capacitance (C) of the process was evaluated. For a CPE with admittance $Q$ and ideality factor $\alpha$, the capacitance is given by:

$$C = \left(\frac{Q}{R^{\alpha-1}}\right)^{1/\alpha} \tag{S1}$$

where $R$ is the resistance parallel to the CPE. The extracted capacitance value for Ag$_8$$T$S$_6$ ($T$ = Si, Ge, Sn) is approximately $10^{-11}$ F, indicating in-grain ionic conduction.[11] The low-frequency tail corresponds to the blocking effect of the electrodes. The high-frequency resistance primarily arises from the combined ionic transport through Ag$_8$$T$S$_6$ and the RbAg$_4$I$_5$ electrode layers. However, due to the thin electrode layer and high ionic conductivity of RbAg$_4$I$_5$, its contribution is negligible. Therefore, to accurately determine the intrinsic ionic conductivity of Ag$_8$$T$S$_6$, the resistance contribution of the electrode material is subtracted from the total resistance. The ionic conductivities of Ag$_8$$T$S$_6$ ($T$ = Si, Ge, Sn) are evaluated across a range of temperatures and exhibited Arrhenius-type behavior (**Figure S16**). All three compositions showed comparable ionic conductivities (**Figure 8b**). The room-temperature ionic conductivity values and the corresponding activation energies for ion transport are summarized in the table below (**Table S12**).

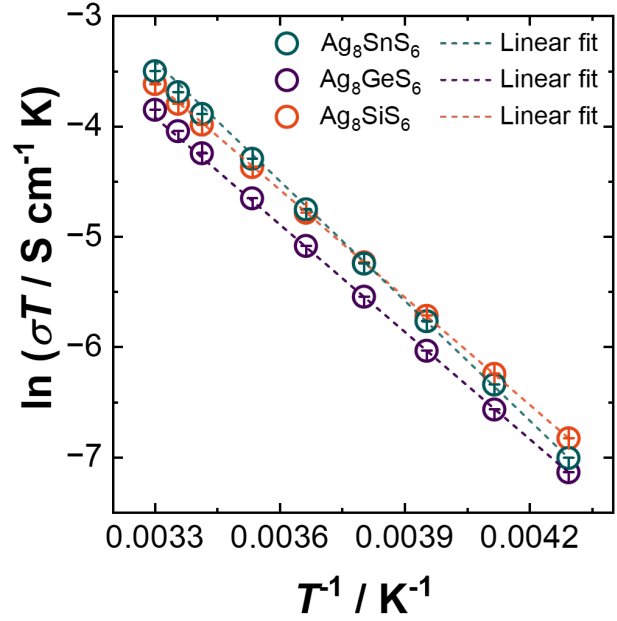

*Figure S16.* Arrhenius plots of measured conductivities of Ag$_8$TS$_6$ (T = Si, Ge, Sn), indicating an activation energy of 0.29 ± 0.2 eV for ion transport.

*Table S12.* The ionic conductivities at 298 K and the obtained activation energy barriers for ion transport in Ag$_8$TS$_6$ (T = Si, Ge, Sn).

| T in Ag$_8$TS$_6$ | $\sigma_{ion}$ at 298 K / mS/cm | $E_a$ / eV |
|---|---|---|
| Si | 0.075 ± 0.008 | 0.27 ± 0.02 |
| Ge | 0.065 ± 0.005 | 0.28 ± 0.01 |
| Sn | 0.081 ± 0.007 | 0.31 ± 0.02 |

**Section S6: Thermal transport – Computational details**

**Sound velocity**

Debye temperature ($\theta$) and frequency ($\omega_D$) can be estimated either from elastic properties calculations or derived from phonon computations. In this work, we used the phonon-based reduced Debye frequency, which corresponds to the acoustic Debye frequency ($\omega_{AC} = N^{-1/3}\omega_D$), where $N$ is the number of atoms in the unit cell. This frequency limit is more suitable for capturing the anharmonicity of acoustic phonons, which are the main contributors to phonon-phonon scattering and thermal transport properties.

**Table S13.** *Debye temperatures obtained through elastic property calculations ($\Theta_{EL}$) and phonon computations, including all phonon modes ($\Theta_{phonon}$) and acoustic modes ($\Theta_{AC}$) compared with Debye temperature derived from experimental measurements $\Theta_{Exp}$.*

| $T$ in Ag$_8$TS$_6$ | $\Theta_{EL}$ (K) | $\Theta_{phonon}$ (K) | $\Theta_{AC}$ (K) | $\Theta_{Exp}$ (K) |
|---|---|---|---|---|
| Si | 170 | 227 | 94 | 163.8±5.5 |
| Ge | 167 | 173 | 82 | 176.0±7.1 |
| Sn (RT) | 160 | 162 | 78 | 163.0±6.3 |
| Sn (LT) | 172 | 158 | 98 | -- |

For all Ag$_8$TS$_6$ ($T$= Si, Ge, and Sn) compounds, we compare sound velocities calculated from two computational approaches (derived from elastic and phonon calculations) with our experimental measurements. For Ag$_8$GeS$_6$ and Ag$_8$SnS$_6$ (RT), both computational and experimental results show no significant differences. In contrast, Ag$_8$SiS$_6$ exhibits more pronounced discrepancies between our calculations and experimental data. This is especially true for the results derived from phonon calculations.

**Table S14.** *Sound velocity ($v_l$, and $v_t$) derived from elastic property calculations. $v_m$ were obtained following Eq S2. $v_{phonon}$ is the sound velocity calculated from phonon calculation and using Eq. 2 from the main text. Here, $v_m^*$ is the experimental mean sound velocity at 300 K (this study).*

| $T$ in Ag$_8$TS$_6$ | $v_l$ (m/s) | $v_t$ (m/s) | $v_m$ (m/s) | $v_{phonon}$ (m/s) | $v_m^*$ (m/s) |
|---|---|---|---|---|---|
| Si | 2982.68 | 1389.15 | 1564.27 | 2087 | 1487±50 |
| Ge | 2960.67 | 1368.95 | 1542.06 | 1601 | 1602±65 |
| Sn (RT) | 2923.28 | 1332.50 | 1501.98 | 1513 | 1501±58 |
| Sn (LT) | 3042.83 | 1417.60 | 1596.28 | 1476 | -- |

$$v_m = \left[\frac{1}{3}\left(\frac{2}{v_t^3} + \frac{1}{v_l^3}\right)\right]^{-1/3} \qquad (S2)$$

**Lattice thermal conductivity in compounds with high anharmonicity**

The diversity in composition and structures, as well as the complexity of the materials involved, make it difficult to determine the lattice thermal conductivity using one specific model. Thermal conductivity models range from simple empirical relationships to complex quantum mechanical calculations. For instance, it is possible to estimate the minimum thermal conductivity using the Cahill[12] and Agne[13] models. Cahill follows Einstein's notion of lattice vibration. The model assumes that the individual oscillators vibrate independently, and that the phonon relaxation time is half the vibration period. Given the inverse correlation between minimum thermal conductivity and speed of sound, a lower minimum thermal conductivity is therefore expected when the speed of sound is low. On the other hand, Agne proposed a diffusion-mediated model in which the phonon density of states is used and it is assumed that all vibrations behave as diffusons with a jump distance equal to a characteristic interatomic distance. Both models can be computed as follows:

$$\kappa_{\min}^{\text{Cahill}} = \frac{1}{2.48} k_{\text{B}} \left(\frac{N\rho N_A}{M}\right)^{2/3} (v_{\text{L}} + 2v_T) \quad (S3) \qquad \kappa_{\min}^{\text{Agne}} = 0.76 n^{2/3} k_{\text{B}} \frac{1}{3} (v_{\text{L}} + 2v_T) \quad (S4)$$

Where $k_{\text{B}}$ and $N_A$ are the Boltzmann and Avogadro constants, $N$ is the number of atoms, $\rho$ is the density, and $v_{\text{L}}$ and $v_T$ are the longitudinal and transversal sound velocities, respectively.

The Slack model, [14–16] as mentioned in the main text, is also an alternative to compute the lattice thermal conductivity of materials. Here, the acoustic modes play an important role in the thermal transport process and the lattice thermal conductivity can be computed as:

$$\kappa_{\text{Slack}} = A \frac{\overline{M} \delta n^{1/3} \Theta_{\text{AC}}^3}{\gamma^2 T} \qquad (S5)$$

Where $\overline{M}$ is the average atomic mass, $V$ is the volume of the unit cell, $\Theta_{\text{AC}}$ is the acoustic Debye temperature, $T$ is the absolute temperature, $k_{\text{B}}$ and $\hbar$ are the Boltzmann and Planck constants, respectively, and $A$ is the Slack coefficient, which is dependent on the anharmonicity of the structure, represented by the Grüneisen parameter.

$$A = \frac{2.436 \times 10^{-8}}{1 - \frac{0.514}{\gamma} + \frac{0.228}{\gamma^2}} \qquad (S6)$$

This model can provide helpful information about lattice thermal conductivity; however, it tends to be generally overestimated compared to experimental data. This discrepancy can be related to the *A* coefficient, but can be effectively corrected by scaling by a factor, as proposed by Qin[17]:

$$A = \frac{0.609 \times 10^{-6}}{1 - \frac{0.514}{\gamma} + \frac{0.228}{\gamma^2}} \tag{S7}$$

In the two-channel model by Simoncellli et al.[18], the heat flux and thermal conductivity are expressed by a matrix, where the diagonal part describes the phonon modes conducting heat as propagating waves known as the phonon-gas channel, and the non-diagonal part represents the phonons conducting energy diffusely, known as the diffuson-channel or "random walk". The sum of those two channels gives the total lattice thermal conductivity.

$$\kappa = \kappa_{\text{ph}} + \kappa_{\text{Diff}} \tag{S8}$$

Studies of similar argyrodite-type materials suggest that the Ag$^+$ vibration has a non-propagating diffuson-like character.[9] To corroborate this, we also compute the thermal conductivity following the recent two-channel model developed by Xia et al [19], which uses the harmonic phonons and considers that each phonon's lifetime is half of its vibration period:

$$\kappa_{\text{Xia}}^{min} = \frac{\pi \hbar^2}{\kappa_B T^2 V N_q} \sum_q \sum_{s,s'} \frac{\left(\omega_q^s + \omega_q^{s'}\right)^2}{2} v_q^{s,s'} \otimes v_q^{s',s} \frac{\omega_q^s n_q^s (n_q^s + 1) + \omega_q^{s'} n_q^{s'} (n_q^{s'} + 1)}{4\pi^2 \left(\omega_q^{s'} - \omega_q^s\right)^2 + \left(\omega_q^s + \omega_q^{s'}\right)^2} \tag{S9}$$

In **Table S15,** we show the lattice thermal conductivity following all models mentioned above. Cahill, Agne and Xia models accurately estimate the minimum lattice thermal conductivity (κmin) for all the argyrodite compounds. These models are valuable, but their applicability is inherently limited to estimating theoretical lower bounds. In addition, the scaled Slack model reduced the overestimation shown in the original Slack model and also agrees with the lower thermal conductivity, especially at 600K.

*Table S15. Comparison of lattice thermal conductivity following different models.*

| | Thermal conductivity ($W/mK$) | | | | | | | |
|---|---|---|---|---|---|---|---|---|
| | | | $\kappa_{Xia}^{min}$ (T=600K) | | | | | |
| $T$ in Ag$_8$$T$S$_6$ | $\kappa_{Cahill}^{min}$ | $\kappa_{Agne}^{min}$ | $\kappa_{diff}$ | $\kappa_{phonons}$ | $\kappa_{Total}^{min}$ | $\kappa_{Slack}$ | $\kappa_{Scaled-Slack}$ | $\kappa_{exp}^{min}$ * |
| Si | 0.428 | 0.269 | 0.260 | 0.010 | 0.271 | 3.26 | 0.82 | 0.411±0.024 |
| Ge | 0.420 | 0.264 | 0.235 | 0.011 | 0.245 | 0.73 | 0.18 | 0.434±0.030 |
| Sn (RT) | 0.402 | 0.253 | 0.222 | 0.011 | 0.233 | 0.91 | 0.23 | 0.391±0.025 |
| Sn (LT) | 0.421 | 0.265 | 0.202 | 0.020 | 0.221 | 2.41 | 0.60 | |

*$\kappa_{exp}^{min}$ minimal lattice thermal conductivity was derived from sound velocity measurements at 300K

In contrast to some of the above approximations, our full ab initio model explicitly incorporates anharmonic effects, enabling the prediction of temperature-dependent lattice thermal conductivity. Before estimating the thermal conductivity with the Grüneisen parameter-based approach, we compare the phonon lifetimes obtained from the analytical model, foundation machine-learned interatomic potentials (MACE-MP-03b), and the Grüneisen parameter-based approach. In **Figure S17a**, at 10K, the analytical model shows that both point-defect and boundary scattering have a strong impact on the phonon lifetimes, and consequently, influence the material's thermal conductivity and phonon transport properties. This observation is consistent with the trend of the phonon channel shown in **Figure 7d** in the main text. Here, MACE-MP-03b and Grüneisen models also exhibit reasonable agreement with the analytical model when point-defect and boundary scattering are subtracted. At 300K (**Figure S17b**), the effect of the point-defect and boundary scattering is less evident from the analytical model, but some differences are still present in both computational approaches. We anticipate that these differences in the life times will also be present in the predicted lattice thermal conductivity, especially at lower temperatures. Nevertheless, at 300K, the agreement between the models improves, which is also reflected in the close values of the calculated lattice thermal conductivity. A complete comparison of the lattice thermal conductivity of Ag$_8$GeS$_6$ is also presented in **Figure S19** and **Table S16**.

Additionally, we investigated the impact of different cutoff frequencies for the average Grüneisen parameter. As shown in **Figure S17b**, focusing on the acoustic modes yields good agreement with the experimental Grüneisen parameter, particularly for the Ag$_8$GeS$_6$ and Ag$_8$SnS$_6$ argyrodites. For Ag$_8$SiS$_6$, a slight deviation is observed, consistent with the differences shown in the Grüneisen parameter plot in **Figure 6** in the manuscript. We employed the average derived from the acoustic Grüneisen parameter in the subsequent calculations of lattice thermal conductivity.

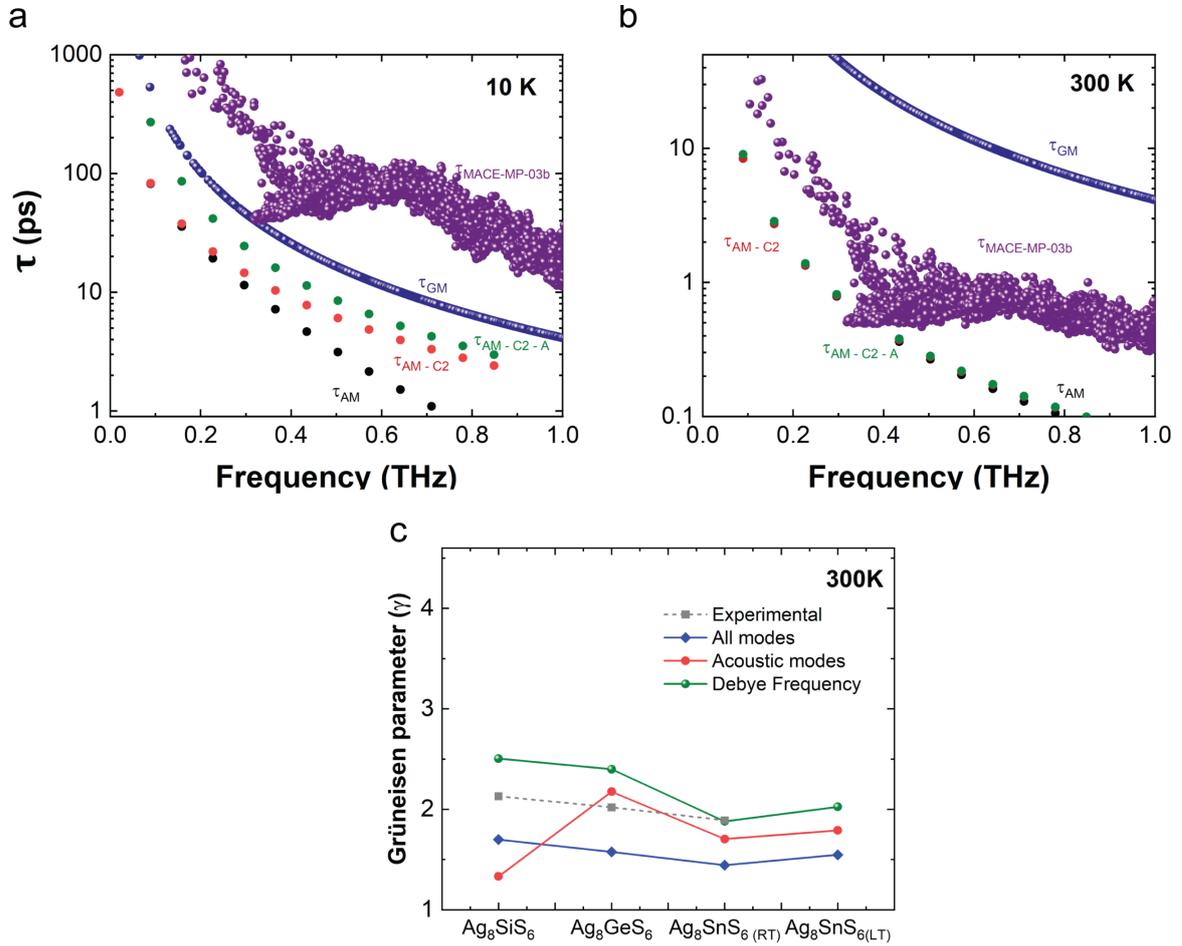

*Figure S12*. a) Phonon lifetimes as a function of the phonon vibrational frequencies for $Ag_8GeS_6$ at 10K. Here, point defects and boundary scattering have a considerable influence on the phonon lifetimes in the analytical model. The foundation model MACE-MP-03b and Grüneisen Model present reasonable agreement with it. b) Phonon lifetimes at 300K for the same models mentioned before. The two computational models agreed well with the phonon lifetimes predicted by the analytical model. c) Average Grüneisen parameter for $Ag_8TS_6$ (T = Si, Ge, Sn), evaluated with different frequency cutoffs. Here, a q-mesh of 10×20×14 was found to be optimal and subsequently used for the lattice thermal conductivity calculations.

Finally, using the Grüneisen model to compute the lattice thermal conductivities shows very good agreement with the experimental results for the sulfide-argyrodite compounds, as illustrated in **Figure S18** and **Figure 7a.** Our computational results show a slight deviation for $Ag_8SiS_6$ from experiment. These deviations are attributed to variations in the mode-dependent Grüneisen parameters.

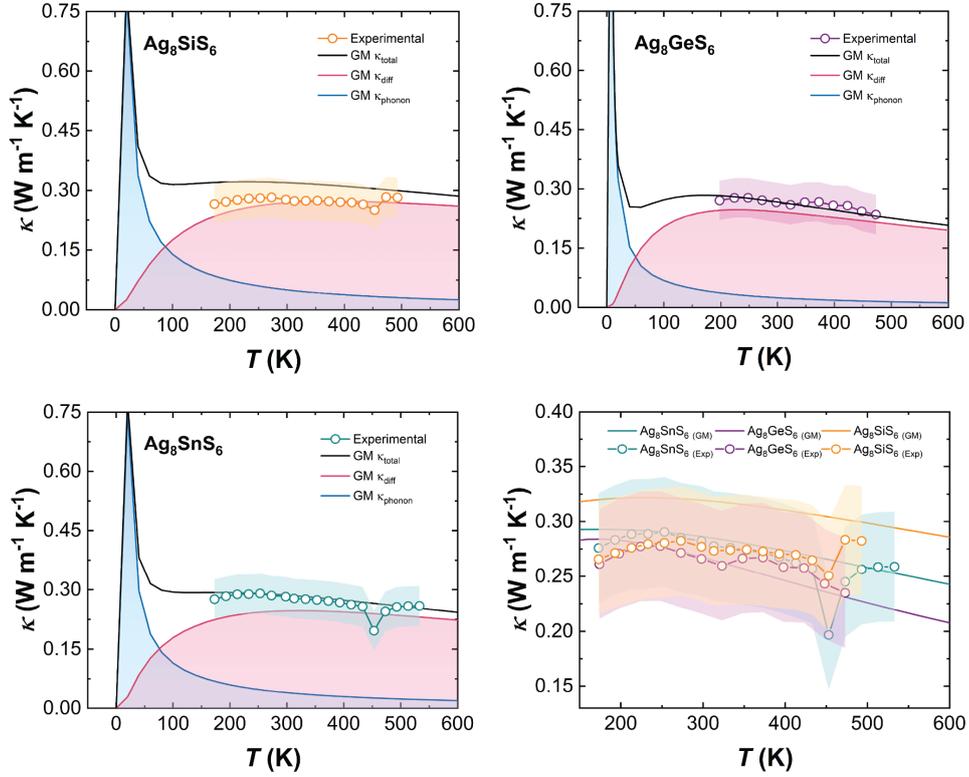

***Figure S13.*** *Two-channel model using our proposed Grüneisen model for the $Ag_8TS_6$ (T = Si, Ge, Sn) compared with our experimental measurements.*

For the $Ag_8GeS_6$ argyrodite, a full comparison of the total lattice thermal conductivity is presented in **Figure S19** and **Table S16-S17.** The two models developed in this work (Grüneisen and ML) show minor differences, and they remain consistent with the experimental results, with particularly strong agreement observed at temperatures above 200 K. Additionally, in **Table S18,** we include the fitting parameters from the analytical model presented in **Figure 7a** in the main text.

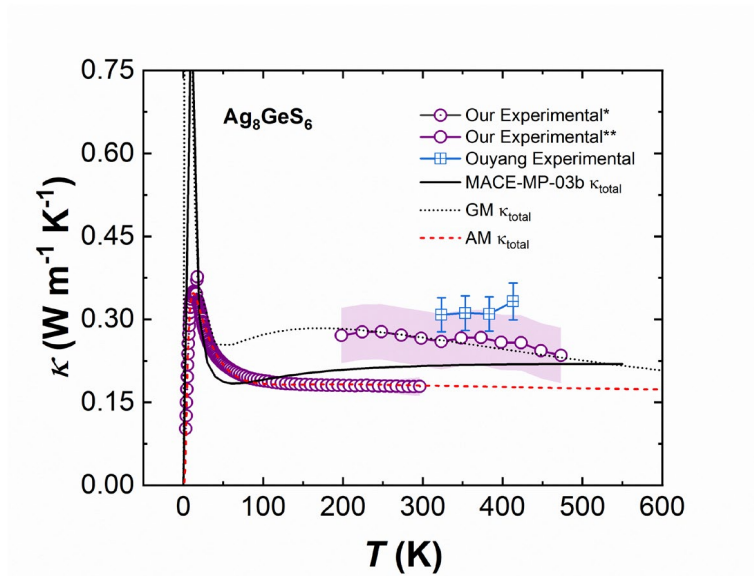

*Figure S19.* Total lattice thermal conductivity comparison for Ag$_8$GeS$_6$: Analytical, Grüneisen, and foundation MACE-MP-30b model versus experimental data (this work and literature).

*Table S16.* Comparison of lattice thermal conductivity of Ag$_8$GeS$_6$ argyrodite obtained from the Grüneisen model, analytical model fitted to low-temperature experimental data from our work, and other experimental measurements (Ouyang and this work).

| Ag$_8$GeS$_6$ | $\kappa_{(300K)}$ (W m$^{-1}$ K$^{-1}$) |
|---|---|
| Mace-MP-03b | 0.216 |
| Full Analytical model | 0.181 |
| Grüneisen model | 0.267 |
| Our Exp** | 0.266 ± 0.050 |
| Ouyang $_{exp}$ | 0.274 ± 0.028 |
| Ouyang $_{comp}$ | 0.312 |

***Table S17.*** *Comparison of the Xia, our Grüneisen model, and experimental results for $Ag_8GeS_6$. The phonon channel is more pronounced in our Grüneisen-based model.*

| | Lattice thermal conductivity ($W/mK$) | | | | | | |
|---|---|---|---|---|---|---|---|
| | $\kappa_{Xia}^{min}$ (T=400K) | | | $\kappa_{GM}$* (T=400K) | | | $\kappa_{Exp}$ (T=433K) |
| $T$ in $Ag_8TS_6$ | $\kappa_{diff}$ | $\kappa_{ph}$ | $\kappa_{Total}^{min}$ | $\kappa_{diff}$ | $\kappa_{ph}$ | $\kappa_{Total}^{min}$ | |
| Si | 0.257 | 0.010 | 0.267 | 0.273 | 0.038 | 0.310 | 0.267 ± 0.05 |
| Ge | 0.232 | 0.010 | 0.243 | 0.227 | 0.019 | 0.246 | 0.252 ± 0.05 |
| Sn (RT) | 0.220 | 0.011 | 0.231 | 0.244 | 0.030 | 0.273 | 0.257 ± 0.05 |
| Sn (LT) | 0.199 | 0.020 | 0.219 | 0.219 | 0.043 | 0.262 | - |

* GM corresponds to the Grüneisen-based model

***Table S18.*** *Fitting parameter extracted from the analytical model of the $Ag_8GeS_6$ argyrodite. $C_1$ represents phonon-phonon scattering, $C_2$ is the point-defect scattering coefficient, A is related to boundary scattering, and P is related to the overlap integral between the linewidths of two proximal phonon modes.*

| | Fitting parameters |
|---|---|
| $C_1$ | $11.801 \pm 0.284 \times 10^{-16}$ s K$^{-1}$ |
| $C_2$ | $15.431 \pm 1.848 \times 10^{-40}$ s$^3$ |
| P | $0.387 \pm 0.007$ |
| A | $0.015 \pm 0.002$ |

**Validation of the Grüneisen model**

To validate our Grüneisen-based model, we computed the lattice thermal conductivity for similar argyrodite materials. Here, $Ag_8GeSe_6$ and $Ag_9GaSe_6$ were considered. First, we started with the phonon dispersion curves and Grüneisen parameter plots for both compounds (**Figure S20**).

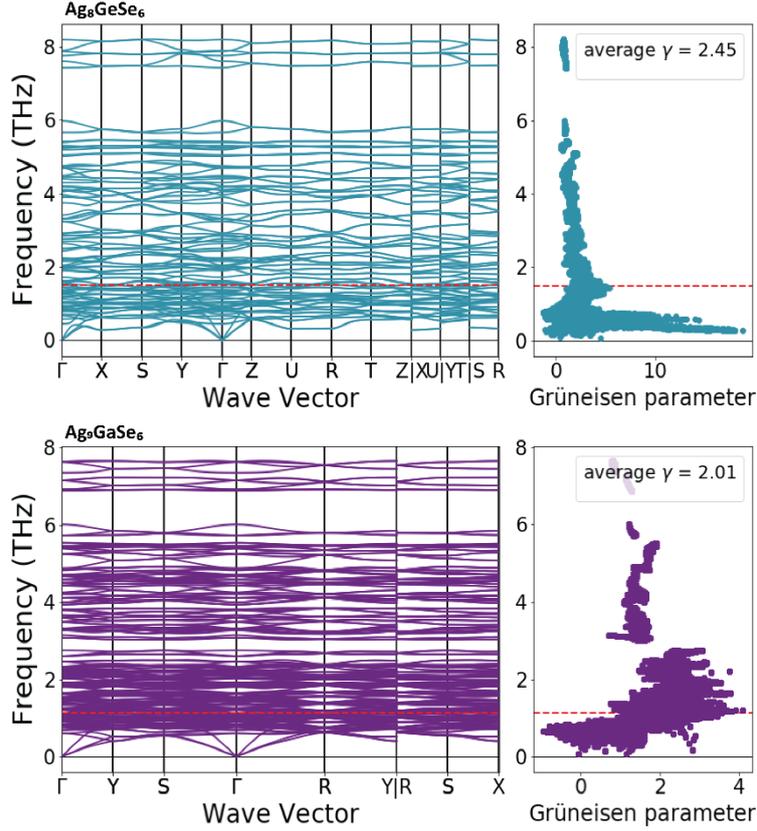

*Figure S20. Phonon band structure and Grüneisen parameter for Ag$_8$GeSe$_6$ and Ag$_9$GaSe$_6$*

The lattice thermal conductivity calculated with our model, which includes only the phonon-phonon scattering, reproduces the experimental result with good enough accuracy, considering the associated experimental uncertainties for both compounds. In **Figure S9 a and b**, we show the two-channel model for Ag$_8$GeSe$_6$. At the low-temperature range, we observe an overestimation in the phonon channel, which influences the total lattice thermal conductivity. However, this is not surprising as we neglect point-defect and microstructure effects in our Grüneisen-based model. In the room- and high-temperature (between 250-600K) range, the thermal conductivities show better agreement with the analytical model and the experimental values reported by Bergnes. Similar results were observed for the Ag$_9$GaSe$_6$. Overall, the total lattice thermal conductivity for both structures are within the expected error margins, supporting the reliability and accuracy of the results.

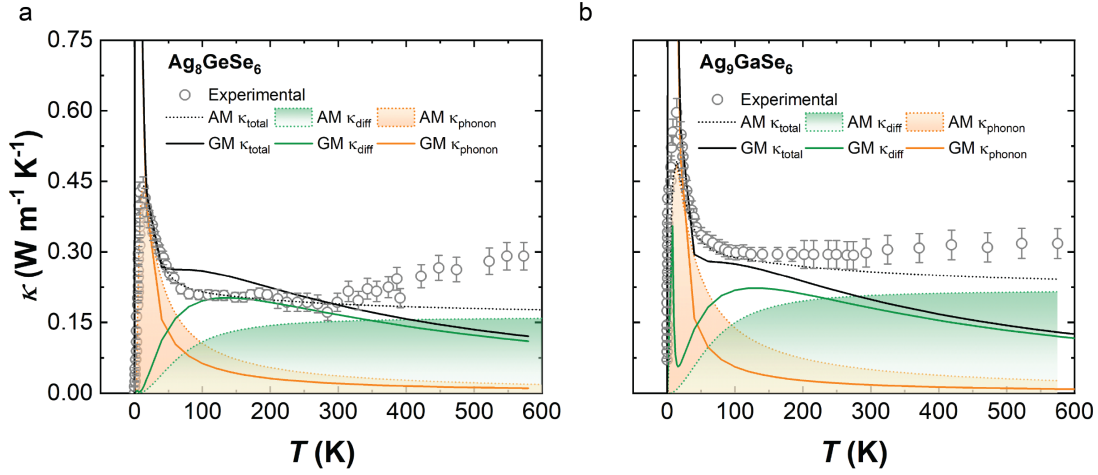

*Figure S21.* Comparison of our proposed Grüneisen-based two-channel model compared with the experimental and analytical model published by Bergnes[20] for both a) $Ag_8GeSe_6$ and b) $Ag_9GaSe_6$ argyrodites.

**Section S7: Possible phase transition**

When performing the quasi-harmonic approximation, one of the expanded volume structures of $Ag_8Sn_1S_6$ shows a new possible phase transition, which, to our knowledge, has not been demonstrated experimentally. The phonon dispersion curve and PDOS, show the softening of the modes at expanded volume (typically connected to a higher temperature), leading to a possible new phase transition or dynamic stability. The new phase has the same space group as the RT structure (P$na2_1$) and we can represent it with the same unit cell. The primitive unit cell that we chose for representation contains 60 atoms, which caused 180 phonon modes. A small expansion in the cell is observed due to the change in the position and coordination environments of the Ag atoms (**Figure S22**). In **Table S18**, we report the different coordination environments. Ag1 and Ag2 change to linear coordination from trigonal planar and tetrahedral, respectively. Another change is observed in Ag6 and Ag8, where the triangular non-coplanar environments change to trigonal planar and tetrahedral coordination environments. The coordination environments were again determined by quantum-chemical bonding analysis with LOBSTER and LobsterPy.

*Table S19.* Atomic positions and inequivalent site fractional coordinates of the possible phase transition of $Ag_8SnS_6$.

| Atom | POSCAR Position | Wyckoff Positions | x | y | z | Coordination Environment |
|---|---|---|---|---|---|---|
| Ag1 | Ag1 | 4a | 0.041 | 0.016 | 0.062 | Linear |
| Ag2 | Ag5 | 4a | 0.026 | 0.142 | 0.324 | Linear |
| Ag3 | Ag9 | 4a | 0.132 | 0.261 | 0.780 | Trigonal Planar |
| Ag4 | Ag13 | 4a | 0.230 | 0.990 | 0.006 | Trigonal Planar |
| Ag5 | Ag17 | 4a | 0.267 | 0.108 | 0.337 | Linear |
| Ag6 | Ag21 | 4a | 0.266 | 0.353 | 0.107 | Trigonal Planar |
| Ag7 | Ag25 | 4a | 0.405 | 0.914 | 0.584 | Linear |
| Ag8 | Ag29 | 4a | 0.444 | 0.044 | 0.435 | Tetrahedral |
| **Sn1** | Sn33 | 4a | 0.130 | 0.712 | 0.761 | Tetrahedral |
| S1 | S37 | 4a | 0.999 | 0.269 | 0.637 | Tetrahedral |
| S2 | S41 | 4a | 0.129 | 0.256 | 0.004 | Trigonal bipyramidal |
| S3 | S45 | 4a | 0.121 | 0.454 | 0.385 | See-saw like |
| S4 | S49 | 4a | 0.248 | 0.207 | 0.629 | Tetrahedral |
| S5 | S53 | 4a | 0.385 | 0.269 | 0.259 | Square pyramidal |
| S6 | S57 | 4a | 0.361 | 0.457 | 0.899 | Triangular non-coplanar |

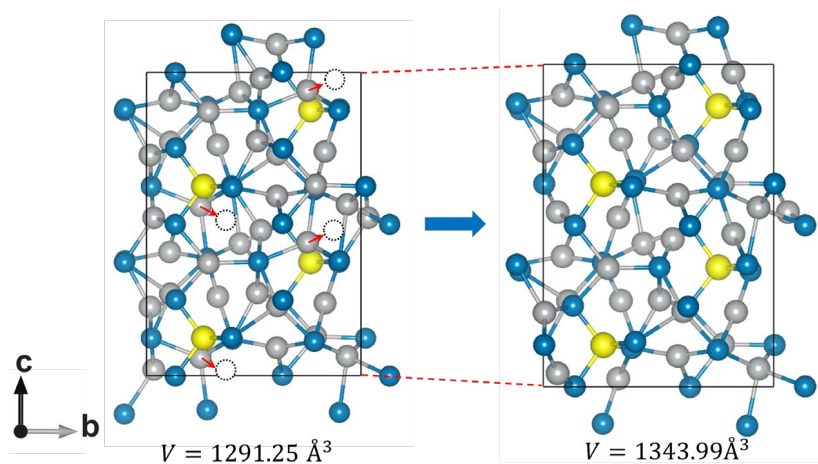

*Figure S22.* Structure change of the possible phase transition. A volume expansion is observed in $Ag_8SnS_6$ argyrodites.

The shape of the phonon dispersion curves is very similar to the room and low-temperature structures. The acoustic modes have a dominating peak in the frequency of 1.6 THz, which originated from the heavy Ag atoms, as can be observed in **Figure S23**. However, the experimental analysis does not show evidence of a new phase transition. This new phase could also be an artefact of the DFT functional or connected to the mobile nature of the Ag atoms.

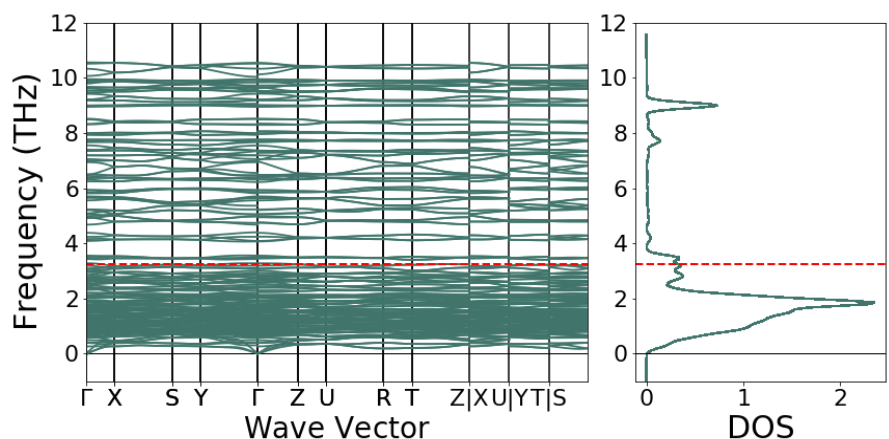

*Figure S23*. Phonon band structure together with phonon density of state for the possible phase transition of Ag$_8$SnS$_6$ argyrodites. The Debye frequency is marked in red.